\documentclass{article}
\usepackage{amsmath,amssymb,amsthm}
\usepackage{tikz}
\usepackage[bottom]{footmisc}
\usepackage{graphicx}
\usepackage[utf8]{inputenc}
\usepackage{mathtools}
\usepackage{pgfplots}
\usepackage{breqn}

\usepackage{subfiles}
\usepackage[justification=centering,font=scriptsize,skip=0pt, belowskip=-1pt,aboveskip=0pt]{caption}
\usepackage{epstopdf}
\usepackage{lipsum}
\usepackage[ruled, vlined, linesnumbered]{algorithm2e}
\SetAlFnt{\small\sffamily}

\SetCommentSty{mycommfont}
\usepackage{color}
\usepackage{subfigure}
\usepackage{pifont}

\usepackage{algorithmic}

\usepackage[english]{babel}
\newtheorem{theorem}{Theorem}[section]

\newcommand{\lightchain}{LightChain\xspace}
\newcommand{\systemCapacity}{n\xspace}
\newcommand{\blockCapacity}{b\xspace}
\newcommand{\transaction}{tx\xspace}
\newcommand{\idSize}{s\xspace}

\newcommand{\validatorThreshold}{\alpha\xspace}
\newcommand{\signatureThreshold}{t\xspace}
\newcommand{\ufp}{q\xspace} 
\newcommand{\adv}{f\xspace} 
\newcommand{\prev}{prev\xspace}
\newcommand{\txnum}{$min\_tx$\xspace}

\newcommand{\storageGain}{66}
\newcommand{\bootstrapGain}{380}

\begin{document}
\title{LightChain: A DHT-based Blockchain for Resource Constrained Environments \footnote{This work has been submitted to the IEEE for possible publication. Copyright may be transferred without notice, after which this version may no longer be accessible.}}

\author{Yahya Hassanzadeh-Nazarabadi, Alptekin K\"{u}p\c{c}\"{u}, and \"{O}znur \"{O}zkasap\\Department of Computer Engineering, Ko\c{c} University, İstanbul, Turkey\\
{\{yhassanzadeh13, akupcu, oozkasap\}}@ku.edu.tr}

\maketitle
\begin{abstract}
As an append-only distributed database, blockchain is utilized in a vast variety of applications including the cryptocurrency and Internet-of-Things (IoT). The existing blockchain solutions have downsides in communication and storage efficiency, convergence to centralization, and consistency problems. In this paper, we propose \textit{\lightchain}, which is the first blockchain architecture that operates over a Distributed Hash Table (DHT) of participating peers. \textit{\lightchain} is a permissionless blockchain that provides addressable blocks and transactions within the network, which makes them efficiently accessible by all the peers. Each block and transaction is replicated within the DHT of peers and is retrieved in an on-demand manner. Hence, peers in \textit{\lightchain} are not required to retrieve or keep the entire blockchain. \textit{\lightchain} is fair as all of the participating peers have a uniform chance of being involved in the consensus regardless of their influence such as hashing power or stake. \textit{\lightchain} provides a deterministic fork-resolving strategy as well as a blacklisting mechanism, and it is secure against colluding adversarial peers attacking the availability and integrity of the system. We provide mathematical analysis and experimental results on scenarios involving $10K$ nodes to demonstrate the security and fairness of \textit{\lightchain}, and show that \textit{\lightchain} is the only existing blockchain that can provide integrity under the corrupted majority power of peers. 
As we experimentally show in this paper, compared to the mainstream blockchains like Bitcoin and Ethereum, \textit{\lightchain} requires around $\storageGain$ times less per node storage, and is around $\bootstrapGain$ times faster on bootstrapping a new node to the system, while each \textit{\lightchain} node is rewarded equally likely for participating in the protocol. 
\end{abstract}

\section{Introduction}
\label{lightchain:sec_intro}
Blockchain \cite{nakamoto2008bitcoin} is an append-only distributed database that provides a partial ordering of blocks among a set of trust-less peers. Each block consists of a set of transactions. In a blockchain, the blocks are connected to each other via immutable links from each block to its previous one and form a chain, which is called the \textit{ledger}. Because they define a partial ordering of blocks without the need of a global synchronized clock, provide a tamper-proof architecture, and establish trust over a trust-less system of independent peers, the blockchain systems are employed in many decentralized applications including the crypto-currencies \cite{nakamoto2008bitcoin}, Internet-of-Things \cite{reyna2018blockchain, khan2018iot}, digital rights management \cite{ma2018blockchain}, big data \cite{mcconaghy2016bigchaindb}, search engines \cite{jiang2017searchain}, fair data exchange \cite{delgado2017fair}, supply-chain management \cite{leng2018research},  P2P cloud storage \cite{hassanzadeh2019decentralized}, and namespace management \cite{kalodner2015empirical}.

A blockchain system is usually modeled as a stack of protocols with at least four layers, from bottom to top are named as \textit{Network, Consensus, Storage, and View} \cite{croman2016scaling}. The layers work interoperably with each other in a pipelined manner i.e., the output of the lower layer is the input to the upper one. The Network layer deals with the dissemination mechanism of the transactions and blocks among the peers of the system. The Consensus layer represents the protocols for block generation decision-making process, which aim at providing an accepted ordering of the blocks among the peers. In other words, all the peers that follow the protocols provided by the Consensus layer are aimed to reach the same state of the generated blocks ordering. The Storage layer provides the read functionality for the peers to read from the blockchain. The View layer represents the most recent state of the participating peers' data considering all the updates on the ledger from the very first to the most recent blocks.
 
\textbf{Existing blockchains' deficiencies: }The existing blockchain solutions have scalability problems in all layers of the blockchain protocol stack. 
To the best of our knowledge, at the Network layer, all the existing blockchains operate on unstructured overlays \cite{nakamoto2008bitcoin, eyal2016bitcoin, nem2018, cryptoeprint:2016:919, kiayias2017ouroboros, king2012ppcoin, bentov2014proof, neo2018, ontology2018, luu2016secure, kokoris2018omniledger, rocket2018snowflake, nikitin2017chainiac, decker2016bitcoin, chepurnoy2016prunable, otte2017trustchain, zamani2018rapidchain, androulaki2018hyperledger, cachin2016architecture}. Such overlays have no deterministic, well-defined, and efficient lookup mechanism to retrieve the address of the peers, the content of the blocks, and the new transactions. Rather, the knowledge of a peer (i.e., other peers, blocks, and transactions) is gained by the epidemic message dissemination among the peers (e.g., broadcasting in Bitcoin \cite{nakamoto2008bitcoin}) with the communication complexity of $O(\systemCapacity)$ to disseminate a new block or transaction, where $\systemCapacity$ is the number of participating peers in the system. In this paper, by the communication complexity we mean the number of the exchanged messages (i.e., the round complexity).

At the Consensus layer, the existing solutions converge to centralization by delegating the block generation decision making to a biased subset of the special peers, e.g., the peers with higher computational power \cite{nakamoto2008bitcoin, decker2016bitcoin, chepurnoy2016prunable, otte2017trustchain}, higher stakes \cite{nem2018, cryptoeprint:2016:919}, or longer activity history in the system \cite{bentov2014proof}. 
Such centralization convergence allows a subset of the peers to leverage the blockchain to their advantage by, for example, performing selfish-mining \cite{eyal2018majority}. The existing blockchains are also prone to the consistency problems that are caused by their probabilistic fork-resolving approach at the Consensus layer, i.e., following the longest chain of the forks as the main chain \cite{nakamoto2008bitcoin}. The probabilistic nature of this fork-resolving approach is due to the volatility of the main chain. 
This threatens the consistency and performance of the system since once the main chain is conquered by another chain, all the blocks that have been already appended to it are considered invaluable \cite{zheng2017overview}. Hence, the probabilistic fork-resolving strategy causes probabilistic finalization on the block generation, i.e., the more blocks are coming after a certain block on the ledger, that chain of blocks gets longer, and with a higher probability that block is being finalized as the main chain's block. Thus, in the existing blockchain solutions, appending a generated block to the ledger does not make it effective unless a number of new blocks come after it \cite{zheng2018blockchain}. 

Having $\blockCapacity$ blocks in the system, the existing blockchains require the Storage layer memory complexity of $O(\blockCapacity)$ by downloading and keeping the entire ledger locally at the peer's storage \cite{croman2016scaling}. In other words, as peers are not able to efficiently lookup any information within the unstructured overlay, they locally store the perceived information and gradually construct a local copy of the entire ledger, which takes $O(\blockCapacity)$ storage complexity. 
Likewise, upon joining the system, during the bootstrapping phase, a new peer needs to verify the entire state of the ledger from the very first block to the most recent one to check the integrity of the ledger \cite{zheng2017overview}. This imposes a time and communication complexity of $O(\blockCapacity)$ at the View layer. Bootstrapping is defined as the process in which a new node constructs its view of the blockchain \cite{croman2016scaling}. 

\textbf{Sharding: }The best existing approach to overcome the mentioned performance and scalability problems of the blockchains is to apply sharding. In the sharding-based approaches \cite{kokoris2018omniledger, luu2016secure, zamani2018rapidchain}, the blockchain system is split into multiple smaller groups of peers, and each group operates in parallel on an independent version of the ledger. Despite its advantage of increasing the speed of the system on processing the transactions in parallel, existing sharding-based blockchains have $O(\systemCapacity)$ communication complexity for processing a single transaction, as well as the best case $O(\frac{\blockCapacity}{\log{\systemCapacity}})$ memory and time complexity at the Storage and View layers, respectively \cite{zamani2018rapidchain}. 

\textbf{Proposed solution: }In this paper, we propose \textit{\lightchain}, which is a permissionless blockchain defined over a Skip Graph-based peer-to-peer (P2P) Distributed Hash Table (DHT) overlay \cite{aspnes2007skip}, with the goal of providing a consistent, communication and storage efficient blockchain architecture with fully decentralized and uniform block generation decision-making. \textit{\lightchain} is permissionless \cite{wust2017you} as it allows every peer to freely join the blockchain system and be considered in the block generation decision-making, which is similar to the well-known blockchains like Bitcoin \cite{nakamoto2008bitcoin} and Ethereum \cite{wood2014ethereum}.
At the Network layer, \textit{\lightchain} operates on top of a Skip Graph that is a DHT-based structured P2P system with a well-defined topology and deterministic lookup strategy for data objects. 
We model each peer, block, and transaction by a Skip Graph node. This idea enables participating peers to make their blocks and transactions addressable and efficiently accessible at the Network layer with the communication complexity of $O(\log{\systemCapacity})$. In other words, each peer, block, and transaction is retrievable by exchanging at most $O(\log{\systemCapacity})$ messages. Additionally, the latest state of each data object is retrievable with the same communication complexity and by querying the Skip Graph overlay directly. By the latest state of a data object, we mean the consideration of all the updates on that data object from the very first block to the most recent one on the ledger. For example, in the cryptocurrency applications where data objects are the peers' balance, the latest state corresponds to the most recent updated value of a peer's balance. This is in contrast to the existing solutions that require the peers to follow the ledger linearly, apply all the updates sequentially, and compute the latest state of a data object. 
As we elaborate in the rest of this paper, we utilize Skip Graph due to its ability to represent each node with two independent identifiers. Nevertheless, \textit{\lightchain} can operate on any DHT with two independent identifiers.

To provide a time and bandwidth efficient consensus approach that is also fair, immutable, and secure, we propose Proof-of-Validation (PoV) as the Consensus layer strategy of \textit{\lightchain}. We say that a consensus approach is fair if each participating peer in the system has a uniform chance of being involved in the consensus regardless of its influence: e.g., processing power, available bandwidth, or stake. In addition, we say that a consensus approach is immutable if none of the (influential) peers in reaching a consensus can legitimately change the consensus at a later time after it is finalized. We say that a consensus approach is secure if the malicious peers are not able to generate and append an illegitimate transaction or block to the ledger. In PoV, the validation of each block is designated to a subset of the peers, which are chosen uniformly for each block based on its hash value (modeled as a random oracle), and are contacted efficiently using the structured Skip Graph overlay. Working in this fashion, \textit{\lightchain} enables improved decentralization of the block generation decision-making and deters the centralization monarchy. By the centralization monarchy, we mean the situation where the majority of block generation decision-makings are under the control of a small subset of special peers e.g., peers with a strong hashing power. \textit{\lightchain} preserves the integrity and consistency of the blockchain in the presence of colluding adversarial peers (e.g., Sybil adversary \cite{douceur2002sybil}) as well as selfish miners \cite{eyal2018majority}, as no peer can contribute to the decision making of two consecutive blocks generation.\footnotemark We discuss these formally in the rest of this paper. \footnotetext{In Section \ref{lightchain:sec_results} we analyze this probability formally, and show that it happens only with a negligible probability in the system's security parameter.}

To improve the consistency of the ledger, \textit{\lightchain} governs a deterministic rule on resolving the forks at the Consensus layer. The main chain is always recognized in a deterministic fashion, and is followed by all the peers. Blocks on the other branches of a fork are discarded by the \textit{\lightchain} peers, i.e., block generation on those branches are rejected by the set of randomly assigned PoV validators, and hence by other peers of the system. This mechanism allows a block to be evaluated and finalized in a deterministic manner as the main chain's block once it is appended to the ledger and one other block comes after it, which is in contrast to the existing solutions that require appending several more subsequent blocks (e.g., around $6$ blocks in Bitcoin \cite{zheng2018blockchain}) to a new block for that block to be considered as a main chain's block.

To establish an efficient Storage layer policy, \textit{\lightchain} enables the peers to access the transactions and blocks in an on-demand basis using the efficient Skip Graph retrievability, rather than requiring them to store the entire ledger locally. Each peer is responsible for keeping a small subset of the randomly chosen blocks and transactions. 
This provides a storage load distribution among the participating peers. To provide better availability of blocks and transactions and tackle malicious peers, \textit{\lightchain} makes several copies of each block and transaction on different peers of the system, which is known as replication \cite{hassanzadeh2018decentralized, hassanzadeh2016awake}. Replication in \textit{\lightchain} is done in a way that it provides at least one copy of each block and transaction accessible at any time in expectation. 

At the View layer, \textit{\lightchain} provides each new peer with a set of randomly chosen peers of the system that are named the \textit{view introducers} of the new peer. The introducers of a new peer are drawn uniformly from the set of participating peers to share their view of the ledger with it. This is done to facilitate the bootstrapping of a new peer joining the system, and enable its immediate participation on the blockchain system without the need to verify the entire blockchain as opposed to the existing solutions. The randomized bootstrapping in \textit{\lightchain} takes $O(\log{\systemCapacity})$ communication complexity and $O(\systemCapacity)$ time complexity. At the end of randomized bootstrapping, a peer obtains the updated view of the most recent state of all the participating peers in the system. However, as we stated earlier, obtaining a particular peer's state in \textit{\lightchain} takes the communication complexity of $O(\log{\systemCapacity})$ and time complexity of $O(1)$, and without the need to track the ledger up to the most recent block. As presented later in this paper, \textit{\lightchain} determines the introducers in a way that the obtained view of a new peer towards the system is consistent with the view of the honest peers.

\textbf{Contributions: }The original contributions of this paper are as follows. 
\begin{itemize}
    \item To the best of our knowledge, this is the first study in the blockchain literature that improves the communication efficiency at the Network layer, the consistency and fairness at the Consensus layer, the memory efficiency at the Storage layer, and provides a more efficient bootstrapping at the View layer, altogether. With this aim, we propose \textit{\lightchain}, which a consistent, and communication and storage efficient permissionless blockchain with fully decentralized and uniform block generation decision-making that operates on top of a Skip Graph-based structured P2P overlay. 
    
    \item \textit{\lightchain} is fair in the sense that each of the participating peers in the system has a uniform chance of being involved in the consensus regardless of its influence: e.g., processing power, available bandwidth, or stake.
    
    \item Having $\systemCapacity$ peers and $\blockCapacity$ blocks in the system, compared to the best existing solutions that require the storage and communication complexity of $O(\frac{\blockCapacity} {\log{\systemCapacity}})$ and $O(\systemCapacity)$ by maintaining many shards, respectively, our proposed \textit{\lightchain} requires  $O(\frac{\blockCapacity}{\systemCapacity})$ storage on each peer, and incurs the communication complexity of $O(\log{\systemCapacity})$ on generating a new block or transaction employing a new blockchain design approach. 
    
    \item In our proposed \textit{\lightchain}, the transactions, blocks, as well as the latest state of the data objects are addressable within the network, and retrievable with the communication complexity of $O(\log{\systemCapacity})$.
    
    \item We provide security definitions for \textit{\lightchain}, and analyze how to set its operational parameters to achieve those security features. 
  
    \item We extended the Skip Graph simulator SkipSim \cite{hassanzadeh2020skipsim} with the blockchain-based simulation scenarios, implemented and simulated the \textit{\lightchain} in extensive simulation scales of $10K$ nodes, and show its performance concerning the security features, in the presence of colluding adversarial nodes.   
 
    \item We also implemented a proof-of-concept version of \textit{\lightchain} node \cite{lightchain-container}, deployed it as an operational \textit{\lightchain} system on Google Cloud Platform, and measured its operational overheads in practice. 

\end{itemize}

The related works are summarized in Section \ref{lightchain:sec_relatedworks}. We describe the preliminaries and our system model in Section \ref{lightchain:sec_preliminaries}. 
Our proposed \textit{\lightchain} is presented in Section \ref{lightchain:sec_solution}. 
We describe the analytical and experimental results in Sections \ref{lightchain:sec_security} and \ref{lightchain:sec_results}, respectively, followed by conclusions in Section \ref{lightchain:sec_conclusion}.

\section{Related Works}
\label{lightchain:sec_relatedworks}
In this section, we survey the existing blockchain solutions based on their contributions to each of the blockchain protocol stack's layers.  

\subsection{Network Layer}
Dissemination of a new transaction or block in the existing blockchains is done via Broadcasting \cite{nakamoto2008bitcoin}, Flooding \cite{chepurnoy2016prunable}, or Gossiping \cite{otte2017trustchain}, which are epidemic disseminations with the communication complexity of $O(\systemCapacity)$ i.e., $O(\systemCapacity)$ message exchanges are required for a single block or transaction to be accessible by every peer of the system. On the other hand, our proposed \textit{\lightchain} applies a communication complexity of $O(\log{\systemCapacity})$ messages to insert a new transaction or block in the Skip Graph overlay, and make it accessible by every peer of the system. Additionally, in our proposed \textit{\lightchain}, not only the blocks, but also the latest state of the data objects are addressable within the network, and retrievable with the communication complexity of $O(\log{\systemCapacity})$. By the latest state, we mean the most recent appearance of that data object in a block on the ledger. By directly retrieving the latest state of a data object, in contrast to the existing blockchains, peers in \textit{\lightchain} are not required to keep searching and retrieving the most recent blocks frequently. Rather, they are able to search and retrieve the latest state of their data objects of interest on demand. For example, in cryptocurrency applications, a peer that is interested only in the latest balance state of another peer, performs a search within the Skip Graph overlay, and finds the latest balance state of the other peer within the blockchain \cite{hassanzadeh2017elats}.  

\subsection{Consensus Layer}
\textbf{Proof-of-Work (PoW):} 
In PoW-based approaches \cite{nakamoto2008bitcoin, jakobsson1999proofs, wood2014ethereum, eyal2016bitcoin}, the block generation is done by tweaking a parameter (i.e., nonce) of the block that makes the hash of it below a predefined difficulty level. 
Considering the hash values are drawn from a uniform distribution (i.e., random oracle model), reaching a hash value below the difficulty level requires a brute force approach over the input range \cite{kraft2016difficulty}.
The block generation decision-making in PoW is heavily correlated with the hash power, which sacrifices the fairness of the system in favor of the nodes with higher hash power \cite{laurie2004proof}. Additionally, PoW is an inefficient consensus solution due to its huge amount of energy consumption \cite{vukolic2015quest} e.g., the power requires to maintain the PoW on Bitcoin network is equal to the Ireland's electrical consumption \cite{o2014bitcoin}.  

\noindent \textbf{Proof-of-Stake (PoS):} In PoS-based approaches, the block generation decision-making is done by the stakeholders \cite{nem2018, cryptoeprint:2016:919, kiayias2017ouroboros, wood2014ethereum, buterin2017casper}. PoS approaches require a synchronized clock \cite{lamport1978implementation} among the peers, which applies an additional $O(\systemCapacity)$ communication complexity. 
PoS approaches also move the system towards the centralized monopoly of the peers with higher stakes and break down fairness and decentralization on block generation \cite{nem2018, wood2014ethereum, buterin2017casper, kiayias2017ouroboros, cryptoeprint:2016:919}.

\noindent \textbf{PoW-PoS Hybrid:}
To provide a balance between the computational inefficiency of PoW and communication inefficiency of PoS, hybrid PoW-PoS approaches are proposed. In PPCoin \cite{king2012ppcoin} the difficulty of PoW is adaptively determined for each peer based on its stake. Similar PoW-PoS hybrid approaches are proposed to combat the spammers in the email systems \cite{dwork1992pricing, laurie2004proof, liu2006proof}. Proof-of-Activity (PoA) \cite{bentov2014proof} is another hybrid approach where peers use PoW over empty blocks to determine the voting committee of the next block uniformly from the set of stake holders.

Compared to the existing PoW and PoS consensus solutions, our proposed Proof-of-Validation (PoV) is the only one that provides fairness, security, and immutability altogether. PoV is fair as it distributes the chance of participating in transactions' and blocks' validation decision-making uniformly among the participating peers, regardless of their influence in the system. In contrast to PoW-based consensus approaches, PoV is secure as malicious peers are not able to generate a validated transaction or block, even with large computational power. In contrast to the PoS-based approaches that are vulnerable to the posterior corruption attack, PoV is immutable and secure against such attacks.  Posterior corruption attack happens when the majority of the committee members of an old block change their decision later on and create a (legitimate) fork from their generated block on the ledger \cite{cryptoeprint:2016:919}. This attack happens especially when the committee members of a block are coming out of stake, and hence do not have anything to lose \cite{poelstra2014distributed}. In our proposed PoV, however, changing even one bit of a transaction or block changes its set of validators entirely. Hence, even the validators of a transaction or block are not able to change its content or fork another history later on.

\noindent \textbf{Byzantine Fault Tolerance (BFT) Consensus and Sharding: } In its classical form, the BFT-based consensus operates on voting nodes. Each node broadcasts its vote to the others, receives their votes, and follows the majority. BFT can tolerate up to $\frac{\systemCapacity}{3}$ of adversarial nodes \cite{pease1980reaching}. Hyperledger \cite{cachin2016architecture, androulaki2018hyperledger}, Ripple \cite{schwartz2014ripple}, and Tendermint \cite{kwon2014tendermint} support BFT-based consensus protocols. For example, in  Ripple, during each epoch, each authority contacts a subset of other authorities as the trusted ones for voting. In sharding-based approaches the system is partitioned into disjoint subsets of peers, e.g., subsets of size $O(\log{\systemCapacity})$ in Rapidchain \cite{zamani2018rapidchain}. Each subset is working on an independent version of the ledger using BFT in an epoch-based manner \cite{luu2016secure, kokoris2018omniledger}. Such epoch-based approaches and BFT apply an additional $O(\systemCapacity)$ communication overhead to the system. NEO \cite{neo2018} is another epoch-based blockchain that aims at resolving forks by Delegated Byzantine Fault Tolerance (dBFT). In dBFT, the participating peers select a set of consensus peers and delegate the block generation decision making to them. Ontology \cite{ontology2018} offers Verifiable Byzantine Fault Tolerance (VBFT), where the consensus peers of the next blocks are selected randomly from the set of stakeholders by applying a random function on the current block. Both dBFT and VBFT require the communication complexity of $O(\systemCapacity)$ for reaching consensus over a block. Snowflake is the consensus layer protocol of Avalanche \cite{rocket2018snowflake}, and acts similarly to our proposed PoV in the sense of randomly chosen peers for validation. However, in contrast to our proposed PoV that engages a small constant number of uniformly chosen peers, Snowflake requires the communal participation of all the online peers for reaching a consensus. Only a small set of trusted super-peers are participated in consensus protocol of BigChainDB \cite{mcconaghy2016bigchaindb}. BigChainDB establishes a variant of Paxos \cite{lamport1998part} consensus among the set of super-peers to elect the one that is responsible for writing to the ledger.

\subsection{Storage Layer}
Having $\blockCapacity$ blocks in the system, existing blockchains like Bitcoin \cite{nakamoto2008bitcoin} and Ethereum \cite{wood2014ethereum}, all require peers to keep an $O(\blockCapacity)$ storage. To moderate this linear storage complexity, Rollerchain \cite{chepurnoy2016prunable} obligates peers to only hold a smaller subset of challenged blocks for generating the new ones. 
The subset, however follows a linear storage complexity in the number of blocks in the system i.e., $O(\blockCapacity)$, which is in contrast to our proposed \textit{\lightchain} that requires $O(\frac{\blockCapacity}{\log{\systemCapacity}})$ storage complexity on each peer. Additionally, Rollerchain lacks the storage load balancing as well as efficient block retrieval features, since blocks are not addressable within the network. Rollerchain also applies a noticeable communication overhead by including the exact copies of the challenged blocks into the newly generated block.
Trustchain \cite{otte2017trustchain} aims to improve the storage load by making each peer to come with its own personal ledger. Each transaction is stored solely on its sender's and receiver's ledgers. In addition to the requirement of a globally synchronized clock and lack of replication, personal ledgers result in $O(\systemCapacity \times \blockCapacity)$ time complexity on generating new transactions. Similar personal ledger approach is also proposed in \cite{harris2018holochain}. Personal ledgers also make the blockchain system not efficiently adaptable to the scenarios where a vast majority of the peers are working on a shared set of data, e.g., distributed database applications. By sharding the system into smaller groups of $O(\log{\systemCapacity})$ peers that operate on disjoint ledgers, Rapidchain \cite{zamani2018rapidchain} requires each peer to keep $O(\frac{\blockCapacity}{\log{\systemCapacity}})$ blocks, but without an efficient retrieval feature. 
BigChainDB \cite{mcconaghy2016bigchaindb} provides a distributed database of super-peers (e.g., Cassandra \cite{cassandra2014apache}) that are the only ones responsible to keep the entire ledger.  Ordinary peers can connect to the super-peers, read the ledger, and propose the transactions. Writing to the database (i.e., ledger) in BigChainDB, however, is only limited to a small set of super-peers that are assumed fully trusted.
Compared to the existing solutions, our proposed \textit{\lightchain} requires $O(\frac{\blockCapacity}{\systemCapacity})$ storage complexity on each peer, and incurs the communication complexity of $O(\log{\systemCapacity})$ on both generation and retrieval of the transactions and blocks, while it presumes a uniform chance for every participating peer to be involved in the block generation decision-making.

\subsection{View Layer}
To the best of our knowledge, there is no existing secure and fast (i.e., $O(1)$ in time and $O(\log{\systemCapacity})$ in communication) bootstrapping approach as we have in our proposed \textit{\lightchain}. Rather, almost all the existing blockchain architectures, including the Bitcoin \cite{nakamoto2008bitcoin} and Ethereum \cite{wood2014ethereum}, require all the peers that are participating in the consensus protocol to construct their view locally by verifying the transactions on the ledger linearly. Having $\blockCapacity$ blocks in the system, this local self-construction of view from scratch takes the time and communication complexity of $O(\blockCapacity)$. To improve the scalability of blockchain and boost up its transaction processing speed, side-chains are proposed as a view-layer solution \cite{back2014enabling}, where a group of peers deviate from the main chain and create their own chain, proceed with their intra-group transactions for a while, and then close the side chain and summarize their turnover by submitting a few transactions into the main chain. Although the side-chains are running faster with significantly fewer peers than the main chain, they are prone to the efficiency problems of the main chain, such as forking. Likewise, as the side-chains grow in number, it is very likely for the sender and receiver of a transaction to reside on two different side-chains, which requires an inter-side-chain transaction. The inter-side-chain transactions should pass through the main chain with possibly two transactions \cite{croman2016scaling} i.e., one deposit from sender's side-chain to the main chain, and one withdrawal from the main chain to the receiver's side-chain. This increases the number of the transactions by a factor of two and acts more of a hurdle for the main chain with advert scalability impact.  

\subsection{Blockchains in relation to the DHTs}
Instead of storing blocks in a linked-list, Skipchain \cite{nikitin2017chainiac} provides a Skip List \cite{pugh1990skip} representation of the blocks in the ledger. Skip List is the centralized analog of Skip Graph. Skipchain enables a peer to search for a specific block within its own local memory in the time complexity of $O(\log{\blockCapacity})$, while it retains the communication complexity of $O(\systemCapacity)$ at the network layer. In the other words, in contrast to our proposed \textit{\lightchain} that enables peers to search for the blocks within the network in a fully decentralized manner, each peer of Skipchain is required to download the entire ledger on its own disk ($O(\blockCapacity)$ storage) and construct the Skip List locally to be able to search for a specific block in logarithmic time. Skipchain is a single-writer blockchain, i.e., only one entity is allowed to write on the blockchain entirely. This limiting assumption is used as appending a new block to Skipchain requires a deterministic knowledge of the meta-data of the immediately subsequent block. 
A blockchain-based decentralized access control management that does not require a trusted third party is proposed in \cite{zyskind2015decentralizing}. The access granted by a user to a service is modeled as a transaction that is encrypted and stored on a DHT \cite{maymounkov2002kademlia}. Users hold the pointers to the encrypted transactions on a blockchain and hence can revoke or change the access grants later on. The blockchain in their proposed solution is mainly utilized as a service without the aim to improve its efficiency, and it is not constructed over the DHT.

Table \ref{lightchain:table_related_works} summarizes a variety of the existing blockchain solutions in comparison to our proposed \textit{\lightchain}.

\begin{table}
\centering
{
    \begin{tabular}{ |l|l|l|l|  }
    \hline
    Strategy & Network &Consensus & Storage\\ 
    \hline
    Bitcoin \cite{nakamoto2008bitcoin} & Broadcast & PoW & Full\\
    BitCoin-NG \cite{eyal2016bitcoin} & Broadcast & PoW & Full\\
    NEM \cite{nem2018} & Broadcast & PoS & Full\\
    Snow White \cite{cryptoeprint:2016:919} & Broadcast  & PoS & Full\\
    Ouroboros \cite{kiayias2017ouroboros} & Broadcast & PoS & Full\\
    PPCoin \cite{king2012ppcoin} & Broadcast  & PoW-PoS & Full\\
    PoA-Bitcoin \cite{bentov2014proof} & Broadcast  & PoW-PoS & Full\\
    NEO \cite{neo2018} & Broadcast & dBFT & Full\\
    Ontology \cite{ontology2018} & Broadcast & VBFT & Full\\
    Elastico \cite{luu2016secure} & Broadcast  & BFT & Full\\
    Ripple \cite{schwartz2014ripple} &  Broadcast & BFT & Full\\
    Tendermint \cite{kwon2014tendermint} & Gossiping & BFT & Full\\
    Hyperledger\cite{androulaki2018hyperledger, cachin2016architecture} & Gossiping & BFT & Full\\
    Omniledger \cite{kokoris2018omniledger} & Gossiping & BFT & Full\\
    Avalanche \cite{rocket2018snowflake} & Gossiping & Snowflake & Full\\
    Skipchain \cite{nikitin2017chainiac} & Gossiping & BFT & Full\\
    PeerCesus \cite{decker2016bitcoin} & Flooding & PoW-PoS & Full\\
    Rollerchain \cite{chepurnoy2016prunable} & Flooding  & PoW & Distributed\\
    Trustchain \cite{otte2017trustchain} & Gossiping  & PoW & Distributed\\
    Rapidchain \cite{zamani2018rapidchain} & Gossiping & BFT & Distributed\\
    BigChainDB \cite{mcconaghy2016bigchaindb} & Broadcast & Paxos & Distributed\\
    \textbf{\textit{\lightchain}} & \textbf{DHT} &
    \textbf{PoV} & \textbf{Distributed}\\
    \hline
    \end{tabular}
}
\caption{A comparison among a variety of the existing blockchain solutions.
We assume that an approach supports distributed storage, if the storage load of blocks and transactions is distributed among all the participating peers in a policy-based manner, e.g., replication. Otherwise, we presume the full storage where peers collect and hold the blocks and transactions entirely on their local storage.}
\label{lightchain:table_related_works}
\end{table}

\section{Preliminaries and System Model}
\label{lightchain:sec_preliminaries}
\subsection{Skip Graph}
Skip Graph \cite{aspnes2007skip} is a DHT-based distributed data structure that consists of nodes. A Skip Graph node is a standalone component with three attributes; a numerical ID, a name ID, and an (IP) address.
The numerical and name IDs are known as the identifiers of the nodes. In order to join a Skip Graph, it is sufficient for each new node to know only one arbitrary node of the Skip Graph, which is called the introducer of that new node. As a result of the join protocol of the Skip Graph that is executed by the new node in a fully decentralized manner, the new node obtains the attributes of $O(\log{\systemCapacity})$ other nodes that are called the neighbors of the new node, where $\systemCapacity$ is the number of Skip Graph nodes. Knowing its neighbors, each node is able to search and find the address of other nodes of Skip Graph that possess a specific numerical ID or name ID, by employing a search for numerical ID \cite{aspnes2007skip,hassanzadeh2015locality}, or a search for name ID \cite{hassanzadeh2016laras} of those nodes, respectively. Both searches are done with the communication complexity of $O(\log{\systemCapacity})$. As the result of the searches, if the targeted numerical ID or name ID of the search is available in the Skip Graph, the (IP) addresses of their corresponding nodes is returned to the search initiator. Otherwise, the (IP) addresses of the nodes with the most similar identifiers to the search target are returned. 

\subsection{Blockchain}
A blockchain is a linked-list of blocks with immutable links from each block to its previous one \cite{croman2016scaling, etemad2015efficient}. By immutable links, we mean that each block points back to the collision-resistant hash value of its previous block on the chain. The immutable links define an order of precedence over the chain of blocks, which implies that the transactions of a certain block are committed subsequent to the transactions of the previous blocks. Due to the immutable links, the blockchain is considered as an append-only database, and updating a block of the ledger by changing its content is not allowed, and considered as an adversarial act. An update on a block changes its hash value and makes the next subsequent block on the ledger not to pointing to this block's hash value anymore, which corresponds to a disconnection on the ledger. To re-establish the connectivity between the updated block and its subsequent block, the pointer on the subsequent block needs to be refreshed with the new hash value of the updated block. This, in turn, changes the hash value of the subsequent block and breaks the ledger from a new point onward (i.e., the subsequent block). Hence, re-establishing the connectivity after an update on a single block requires refreshing the hash pointers on all the subsequent blocks. In the existing blockchains, re-establishing the connections after an update on a block is correlated with a success probability, e.g., solving a computationally hard problem \cite{nakamoto2008bitcoin} or getting the consent of a specific subset of peers \cite{buterin2017casper}. This correlation makes re-establishing the connectivity of ledger upon changing the content of even a single block a computationally hard problem due to the collision-resistance of the hash functions.

\subsection{Notations}In this paper, we call the last block that is appended to the blockchain as the \textit{current tail} of the blockchain, which is also the tail of the ledger. The first block of a blockchain is known as the \textit{Genesis} block, which is also the head of the linked-list of the ledger. We also define the \textit{previous} relationship as the immutable links from each block to its previous block on the ledger. Blockchain defines a partial ordering of the blocks on the ledger based on the previous relationship. We say that block $blk1$ is the \textit{immediate predecessor} of the $blk2$, if $blk2$ points back to the hash value of $blk1$ as its previous block on the ledger. In this situation, $blk2$ is the \textit{immediate successor} of $blk1$. In this paper, we consider that a block is \textit{committed} to the blockchain if it is being written by the consensus layer protocol of the blockchain to its storage, i.e., the block passes the defined consensus verification and is being appended to the tail of the ledger. We denote the system's security parameter and the system's identifier size by $\lambda$ and $\idSize$, respectively. Also in this paper, we denote the hash function $H:\{0,1\}^{*} \rightarrow \{0,1\}^{\idSize}$ as a random oracle.

\subsection{System Model}
In our system model, each peer corresponds to a device connected to the Internet (e.g., a laptop, smartphone, smart TV) that executes an instance of the \textit{\lightchain} protocol. As detailed in Section \ref{lightchain:sec_solution}, a Skip Graph overlay of peers is constructed by representing each peer as a Skip Graph node. 
We assume that each participating peer joins the Skip Graph overlay using the Skip Graph join protocol in a fully decentralized manner and by knowing one peer of the system \cite{aspnes2007skip}. 
Both identifiers (i.e., name ID and numerical ID) of peers are the hash value of their public key using a collision-resistant hash function. Following this convention, in this paper, we refer to a peer by its identifier, which corresponds to its name/numerical ID. 
We consider the system under churn \cite{stutzbach2006understanding}, i.e., the participating peers are dynamic between offline and online states. We assume the existence of a churn stabilization strategy \cite{jacob2014skip+, hassanzadeh2019interlaced} that preserves the connectivity of the Skip Graph overlay under churn. 
We denote the \textit{System Capacity} by $\systemCapacity$, and define it as the maximum number of registered peers in the system, i.e., $\systemCapacity = O(2^{\idSize})$. We consider all the participating peers as probabilistic Turing machines that run in a polynomial time in the security parameter of the system i.e., their running time is $O(\lambda^{c})$ for some constant $c > 0$. We make this assumption essentially for the reason that participating peers should be able to execute $O(\systemCapacity)$ cost protocols. Following this assumption, $\systemCapacity$ is a polynomial in $\lambda$, which results in $\idSize << \lambda$.   
Similarly, we denote the \textit{Block Capacity} by $\blockCapacity$, and define it as the maximum number of the generated blocks in the system. Similarly, we also consider $\blockCapacity$ as a polynomial in $\lambda$, which results in $\frac{\blockCapacity}{\systemCapacity}$ to be a polynomial in the security parameter of the system. Regarding the synchronization, we assume the system is partially synchronous \cite{tanenbaum2007distributed} meaning that most of the time the process execution speeds and message-delivery times are bounded. To handle the times that the system goes asynchronous where no such bounds are assumed, processes use a timeout to conclude that another process has crashed.

In our system model, we assume that each peer is participating in the blockchain by a set of \textit{assets} as well as a \textit{balance}. The assets set corresponds to the data that the peer initially registers on the blockchain via a transaction, and is able to update it later on by generating new transactions. The balance of a peer is used to cover its transaction generation fees. Although the assets and balance are the same in cryptocurrency applications, nevertheless, we consider them as two distinct attributes of each participating peer in general form, considering other potential applications such as distributed databases. We consider a transaction as a state transition of the assets of the transaction's owner. View of a participating peer in our system model towards the blockchain is a table of $(numID\,, lastblk\,, state\,, balance)$ tuples. Each tuple represents the view of the peer with respect to another peer of the system with the numerical ID of $numID$. The $lastblk$ represents the hash value of the last committed block to the blockchain that contains the most recent transaction of that peer. The view of the associated peer with respect to the current state of the assets of another peer and its remaining balance are represented by $state$ and $balance$, respectively. By the current state, we mean the most recent values of the assets of the peer considering all the generated transactions by that peer from the Genesis block up to the current tail of the blockchain. 

\subsection{Adversarial Model }
We define the availability of the blockchain as the blocks being accessible in a timely fashion \cite{goodrich2011introduction}. We define the integrity of the blockchain as the property that views of the peers towards the blockchain are not being changed, except by appending a new block to the current tail of the blockchain solely by the peers that are included in the consensus protocol.
We assume the existence of a Sybil adversarial party \cite{douceur2002sybil} that adaptively takes control over a fraction $\adv$ of peers in the system. We define the honest peers as the ones that follow the \textit{\lightchain} protocol, and the adversarial peers as the ones that deliberately deviate from the \textit{\lightchain} protocol collectively at arbitrary points. Adversarial peers aim to jointly attack the availability and integrity of the system. 

\subsection{Authenticated Search}
We assume that the search queries over the Skip Graph overlay are authenticated by an authentication mechanism in the presence of the described adversarial peers \cite{boshrooyeh2017guard, taheri2020proof}. By the authenticated searches, we mean that the validity of the search results is publicly verifiable through a search proof that is generated by the signing keys of the participating peers on the search path. The search proof also contains the attributes of the peers on the search query path (e.g., identifier and (IP) address) with the last node on the search path considered as the search result. We assume the success chance of adversary on breaking the authenticated search mechanism and forging a search proof is limited to some negligible function $\epsilon(\lambda)$ (e.g., $\epsilon(\lambda) = 2^{-\lambda}$).

\section{\lightchain: A Permissionless Blockchain over Skip Graph}
\label{lightchain:sec_solution}
\subsection{\textbf{Overview}}
The \textit{\lightchain} protocol is executed independently by each participating peer. In \textit{\lightchain}, we employ a Skip Graph DHT overlay to establish a blockchain.
The peers, as well as the transactions and blocks, are indexed as Skip Graph nodes. Each peer invokes the insertion algorithm of Skip Graph \cite{aspnes2007skip} using its own identifiers and (IP) address and joins the system. Both identifiers of a peer (i.e., its name ID and numerical ID) are hash value of its public key (i.e., verification key). 
As a result of joining the Skip Graph overlay, each peer knows logarithmically other peers, which enables it to efficiently search for any other peer of the system with the communication complexity of $O(\log{\systemCapacity})$. Upon joining the Skip Graph overlay, the peer creates its view of the blockchain using \textit{\lightchain}'s randomized bootstrapping feature without the need to download and process the entire ledger. 

In \textit{\lightchain}, a transaction represents a state transition of the assets of a peer, which is denoted by the \textit{owner} peer of that transaction. For example, in the cryptocurrency applications, the asset of a peer is its monetary wealth, and a transaction models a monetary remittance, which represents the state transition of the monetary wealth of the owner affected by the remittance. The owner peer casts the state transition into a transaction, computes the identifiers of validators, searches for the validators over Skip Graph overlay, and asks them to validate its transaction. In order to be validated, each transaction needs to be signed by a system-wide constant number of validators, where their identifiers are chosen randomly for each transaction to ensure security. In addition to security, the idea of validating transactions makes participating nodes in the block generation needless of going through the validation of individual transactions. 

Once the transaction gets validated, the owner inserts it as a node into the Skip Graph overlay, which makes it searchable and accessible by any other peer. The insertion of the transaction is done by invoking the insertion protocol of Skip Graph using the transaction's identifiers but the (IP) address of the owner peer itself. As we explain later, the identifiers of a transaction are related to its hash value. The Skip Graph peers route the messages on behalf of the transactions they hold. This idea is similar to the other existing DHTs like Chord \cite{stoica2001chord} and Pastry \cite{rowstron2001pastry}. This feature enables \textit{\lightchain} peers to search and find the new transactions. Upon finding new validated transactions, each peer is able to cast them into blocks, go through the validation procedure (similar to the transactions' case), and insert the validated block into the Skip Graph overlay. Each transaction's owner then removes its transaction node from the overlay once it is successfully included in a validated block (for the sake of efficiency). The idea of representing each transaction and block by a Skip Graph node results in any search for the peer or the transactions and blocks that it holds to be routed to the peer's (IP) address, rendering them accessible by every other peer in a fully decentralized manner. Hence, in \textit{\lightchain}'s Skip Graph overlay, there exist three types of nodes: peers, transactions, and blocks. In other words,  
the Skip Graph overlay acts as a distributed database of the transactions and blocks that are owned by peers, which enables each peer to efficiently search for any transaction or block with the communication complexity of $O(\log{\systemCapacity})$. The \textit{previous} relationship of blocks stored in a distributed manner on distinct peers defines a blockchain. 
By making the blocks and transactions efficiently retrievable by search, the participating peers are not required to keep or download the entire ledger. In \textit{\lightchain}, each block or transaction is replicated by its owner and validators to support availability, accessibility, and fault tolerance. By means of searchable blocks and transactions as well as replication, in \textit{\lightchain} we introduce the idea of distributed storage layer for the blockchain where participating peers in the consensus only need to keep and maintain a subset of the blocks, and not the ledger entirely. In the rest of this section, unless stated otherwise, by the term node, we mean a peer.  

\begin{figure} 
\centering
\includegraphics[scale = 0.3]{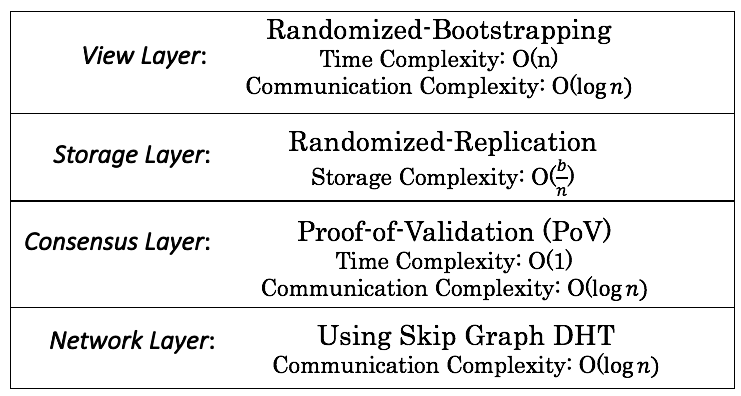}
\caption{\textit{\lightchain}'s protocol stack and its contributions to each layer of the blockchain architecture. The reported asymptotic complexities for the Network layer are per transaction or block, and for other layers are per node.} 
\label{lightchain:fig_lightchain_protocol_stack}
\end{figure}

As an incentive mechanism, \textit{\lightchain} employs a monetary balance for each participating peer to exchange with other peers and cover the operational fees of appending data to the blockchain \cite{nakamoto2008bitcoin}. \textit{\lightchain} rewards the peers' contribution on maintaining the connectivity of the system, providing validation service, and generating blocks. Moreover, \textit{\lightchain} encourages honest peers to audit other peers, by rewarding the detection and report of adversarial acts. Malicious behavior is penalized by \textit{\lightchain} upon detection, and the adversarial peers are blacklisted and gradually isolated from the system. 
Figure \ref{lightchain:fig_lightchain_protocol_stack} summarizes the \textit{\lightchain}'s contributions to each layer of the blockchain architecture.

\subsection{\textbf{Structure of Transactions and Blocks}}
A \textit{\lightchain} transaction, $tx$, is represented by a $(\prev, \, owner, \, cont, \, search\_proof, \, h, \, \sigma)$ tuple, where $\prev$ is the hash value of a committed block to the blockchain. We use the $\prev$ pointer for each transaction $tx$ to define an order of precedence between $tx$ and all the blocks and transactions in the blockchain without the need of any synchronized clock. The block that is referred by $\prev$ takes precedence over $tx$. All the transactions included in the $\prev$ block are assumed to be committed before $tx$ in the essence of time. Following the same convention, all the blocks and transactions that precede $prev$, also precede $tx$. The \textit{owner} represents the identifier of the owner node in the Skip Graph overlay that generates the transaction $tx$. Equating the name ID and numerical ID of the peers with the hash value of their public key, $owner$ refers to either of the name ID or numerical ID of the owner peer. The $cont$ field of a transaction denotes the state transition of the assets of the owner node. The contribution is a general term that covers a vast variety of the blockchain applications that \textit{\lightchain} is applicable on. For example, in cryptocurrency applications, the state of peers corresponds to their wealth, and a transaction represents a monetary remittance between two peers. In such applications, $cont$ includes the remittance value as well as the identifier of the receiver peer, to whom the transaction owner aims to transfer the fund. 
The $search\_proof$ field of a transaction is the authenticated proof of searches over the peers of the Skip Graph overlay to find the validators of the transaction $tx$, as explained before. The $h$ field of the transaction $tx$ is the hash value of the transaction, which is computed as shown by Equation \ref{lightchain:eq_hash_tx}. The $\sigma$ field of the transaction $tx$ contains the signatures of both the owner as well the validators on its hash value $h$. The owner's signature is for the sake of authenticity, and to prevent adversarial peers from masquerading as honest peers and submitting a transaction on behalf of them. The validators' signature is a part of \textit{\lightchain}'s consensus strategy, and is explained within our proposed Proof-of-Validation consensus approach.
\begin{equation}
    h = H(prev||owner||cont||search\_proof) \label{lightchain:eq_hash_tx}
\end{equation}
A \textit{\lightchain} block $blk$ is defined by a $(prev, \, owner, \, \mathcal{S}, \, search\_proof, \, h, \, \sigma)$ tuple, which is similar to the transaction structure of \textit{\lightchain} except that $\mathcal{S}$ represents the set of all the transactions that are included in the block $blk$. The $h$ field of block $blk$ is its hash value, which is computed as shown by Equation \ref{lightchain:eq_hash_blk}. The $\sigma$ field contains the signatures of both the block's owner as well as the block's validators on its hash value (i.e., $h$).  
\begin{equation}
    h = H(prev||owner||\mathcal{S}||search\_proof) \label{lightchain:eq_hash_blk}
\end{equation}

\subsection{\textbf{Network Layer: Skip Graph overlay of peers, transactions, and blocks}}
In our proposed \textit{\lightchain}, we represent each peer, transaction, and block by a Skip Graph node. This way, all the peers, transactions, and blocks are addressable within the network. In other words, participating nodes (i.e., peers) in \textit{\lightchain} exploit the Skip Graph overlay to search for each other, as well as each others' blocks and transactions. Both the numerical ID and name ID of the peers are the hash value of their public key using a collision-resistant hash function. 
As in a Skip Graph, nodes' identifiers define the connectivity; hence, considering the hash function as a random oracle results in the uniform placement of peers in Skip Graph overlay, which limits the adversarial power on tweaking the Skip Graph topology for its advantage.

The numerical ID of a transaction or a block in the Skip Graph overlay is its hash value (i.e., $h$). The name ID of a transaction or a block is its corresponding \textit{\prev} field value. This regulation enables peers to traverse the \textit{\lightchain}'s ledger in both forward and backward directions. Following this convention, in \textit{\lightchain}, having a block with numerical ID (i.e., the hash value) of $h$ and previous pointer value of \textit{prev}, the (IP) address of the peers that hold the immediate predecessor block are obtained by performing a search for numerical ID of \textit{\prev} in the Skip Graph overlay. Similarly, the (IP) address of the peers holding the immediate successor transaction(s) or block(s) in the blockchain are obtainable by performing a search for name ID of $h$ over the Skip Graph overlay. This follows the fact that all the immediate successors of a block have $h$ as their name ID.
This feature of the \textit{\lightchain} enables the peers to efficiently update their view towards the tail of the blockchain by performing a search for the name ID of their local tail. The search returns all the blocks that are appended subsequently to their local tail, as well as all the new validated transactions that are waiting to be included in blocks. Additionally, using this feature, a peer does not need to store the entire blockchain locally. Rather, having only a single block of the ledger enables the peer to efficiently retrieve the predecessor and successor blocks to it with the communication complexity of $O(\log{\systemCapacity})$. 

\begin{figure*}[!]
\centering
\includegraphics[scale = 0.4]{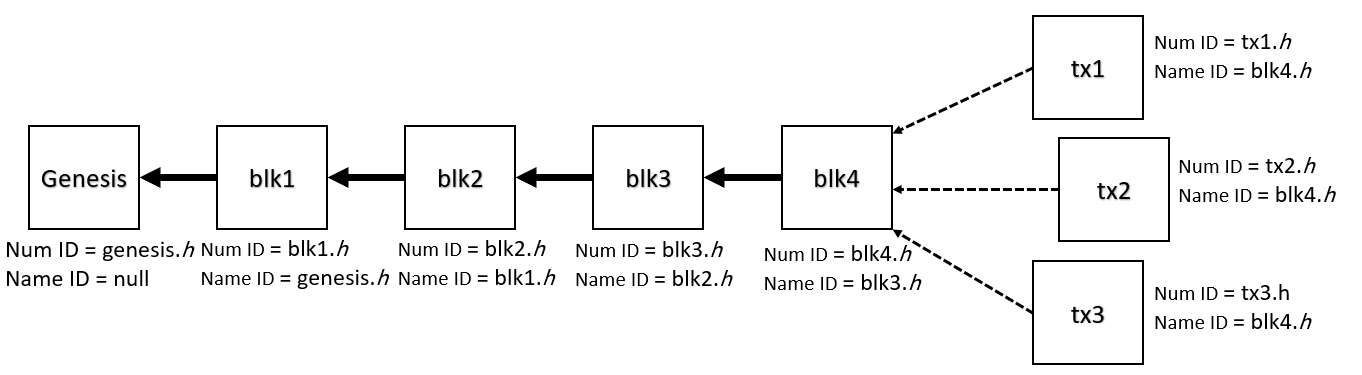}
\caption{The \textit{\lightchain} regulation on name IDs and numerical IDs. Numerical ID (i.e., Num ID) of a block or transaction is its hash value, and name ID is its corresponding \textit{prev} value.} 
\label{figure:lightchain_chain-tx}
\end{figure*}

Figure \ref{figure:lightchain_chain-tx} illustrates this convention of \textit{\lightchain}, where a peer that only has $blk2$ is able to efficiently retrieve its immediate predecessor (i.e., $blk1$) by searching for the numerical ID \cite{aspnes2007skip} of its \textit{\prev} value (i.e., $blk1.h = blk2.prev$) in a fully decentralized manner. The search is responded by the owner \footnotemark of $blk1$ with its (IP) address, and hence the predecessor of $blk2$ (i.e., $blk1$) is retrievable efficiently by directly contacting its owner. Similarly, the peer that only possesses $blk2$ is able to perform a search for name ID \cite{hassanzadeh2016laras} over its hash value (i.e., $blk2.h$) to retrieve the immediate successor block that comes after $blk2$. As the result of the search for name ID of $blk2.h$, owner of $blk3$ responds to the search initiator peer with its (IP) address, and $blk3$ is retrievable efficiently by directly contacting its owner. In the case where a single block has several successor blocks, the search initiator receives a response from each of the immediate successor block's owners. In the example of Figure \ref{figure:lightchain_chain-tx}, considering $blk4$ as the current tail of the blockchain, as discussed later in this section, the newly generated transactions that succeed $blk4$ (i.e., $tx1$, $tx2$, and $tx3$) are efficiently retrievable by performing a search for the name ID using $blk4.h$. 

\footnotetext{Considering the replication of blocks, the search is responded by the either the owner, or one of the replicas. We introduce the replication of blocks and transactions later in this section.}

\subsection{\textbf{Consensus Layer: Proof-of-Validation (PoV), fair, efficient, immutable, and secure consensus}}
Proof-of-Validation (PoV) is our proposed consensus approach of \textit{\lightchain}, and is employed to validate the generated transactions and blocks. Once a transaction or block is validated by PoV, it is considered legitimate by all the participating peers. PoV is fair as each participating peer in the system has a uniform chance of being involved in the consensus regardless of its influence. PoV is efficient as it requires only $O(\log{\systemCapacity})$ communication complexity for validating a single transaction or block. PoV is immutable as none of the influential peers in reaching a consensus can legitimately change the consensus at a later time after it is finalized. Finally, PoV is secure as malicious peers are not able to commit an invalid transaction or block to the blockchain. We analyze the security and immutability of PoV in Section \ref{lightchain:sec_results}. 
A transaction or block is considered as validated once it successfully passes the PoV consensus. Note that a validated transaction's contribution is not considered effective and authoritative unless it is included in a validated block that is committed to the blockchain. To validate each transaction or block, PoV provides a set of randomly chosen validators for the sake of evaluation as detailed in the followings.

\label{lightchain:tx_validation}
\subsubsection{Transaction Generation and Validation}
PoV considers a transaction as valid if its hash value $h$ is signed by $\signatureThreshold$ (randomly chosen) validators, where $\signatureThreshold$ is a constant protocol parameter, which is called the \textit{Signatures Threshold}. 
For a transaction $\transaction$, the numerical ID of each validator is chosen uniformly as shown by Equation \ref{lightchain:eq_validator}, where $v_{i}$ is the numerical ID of the $i^{th}$ validator in the Skip Graph overlay. 
In order to provide security, efficiency, and availability for the system, \textit{\lightchain} only allows a transaction's owner to iterate $i$ over $[1,\validatorThreshold]$, where $\validatorThreshold$ is another constant protocol parameter, which is called the \textit{Validators Threshold}. We formally discuss this in Section \ref{lightchain:sec_results}, and develop a formulation for deciding on the proper values of the \textit{Signatures Threshold} and \textit{Validators Threshold} considering the security, efficiency, and availability of system. 
\begin{equation}
    v_{i} = H(\transaction.\prev||\transaction.owner||\transaction.cont||i) \label{lightchain:eq_validator}
\end{equation}
\noindent The transaction's owner then conducts a search for numerical ID of the validator (i.e., $v_{i}$) within the Skip Graph overlay. If there exists a peer with the numerical ID of $v_{i}$ in the overlay, the owner receives its (IP) address. Otherwise, it receives the (IP) address of the peer with the largest available numerical ID that is less than $v_{i}$. Both cases are supported with an authenticated search proof that is generated by the Skip Graph peers on the search path, and is delivered to the owner. The authenticated proof of the search for numerical ID of the $i^{th}$ validator is denoted by $search\_proof_{i}$, which also contains all the (IP) addresses and identifiers of the Skip Graph peers on the search path. The last peer on the search path of $v_{i}$ is designated as the $i^{th}$ validator. The transaction's owner then adds the authenticated search proof for all the validators to the transaction, computes its hash value $h$ as specified by Equation \ref{lightchain:eq_hash_tx}, signs the hash value, and appends her signature to $\sigma$. The transaction's owner then contacts the validator asking for the validation of the $tx$. During the validation, the validators evaluate the soundness, correctness, and authenticity of $tx$, as well as the balance compliance of the its owner to cover the fees. As the validation result for $tx$, the transaction owner either receives a signature over $h$ or $\bot$ from each contacted validator.

\textbf{Soundness:} A transaction $tx$ is said to be sound if it does not precede the latest transaction of the transaction's owner on the blockchain. By not preceding the latest transaction of the same owner, we mean its $\prev$ should point to the hash value of a validated and committed block on the ledger with no transaction of the transaction's owner in any of the subsequent blocks. In other words, soundness requires the newly generated $tx$ transaction to succeed all of the previously registered transactions of its owner on the blockchain. This is both to counter double-spending from the same set of assets, as well as to make the validation of a transaction a one-time operation, i.e., the owner of a validated $tx$ transaction is able to append it to the blockchain as long as it does not generate any new transaction on the blockchain that precedes $tx$ based on $\prev$.
Considering the soundness, at most one of the concurrently generated and validated transactions of a peer has the chance to be included into a new block. As once one of its transactions is included in a block, the others become unsound, cannot be included in the same block or further blocks, and should go over re-validation. Therefore, in addition to prevent double spending, soundness provides a uniform chance for the transaction generators to include their transaction into each new block. We elaborate more on this criteria when we discuss block validation. 

\textbf{Correctness:} For a transaction $tx$ to be correct, its contribution field (i.e., $cont$) should represent a valid state transition of the owner's assets. The compliance metric is application dependent. For example, in cryptocurrency applications, for a transaction to be correct, the owner's account should have enough balance to cover the remittance fee (i.e., the contribution). 

\textbf{Authenticity:} The evaluation of authenticity is done by checking the correctness of $h$ based on Equation \ref{lightchain:eq_hash_tx}, verifying $\sigma$ for the inclusion of a valid signature of the transaction's owner over $h$, 
and verifying $search\_proof$ for all the validators of $tx$. The validator rejects the validation of $tx$ as unauthenticated if any of these conditions is not satisfied. 

\textbf{Balance Compliance:} As an incentive mechanism to participate in the validation, \textit{\lightchain} considers a validation fee in the favor of the $\signatureThreshold$ validators of the transaction $tx$ that sign its hash value and make it validated. Also, \textit{\lightchain} considers a routing fee in the favor of all the Skip Graph peers that participate in finding the transaction's validators, i.e., the peers that their identifiers are included on the search path of every validator according to the $search\_proof$, excluding the validator and the owner itself. A transaction $tx$ passes the balance compliance part of validation if its owner has enough balance to cover the validation and routing fees. The balance compliance validation is done based on the view of the validator towards the blockchain. Both the routing and validation fees are fixed-value protocol parameters, and are the incentive mechanism for the peers to perform the routing and validation honestly \cite{nakamoto2008bitcoin, eyal2016bitcoin, nem2018}. The fees also prevent Sybil adversarial peers on indefinitely generating transactions by circulating the adversarial balance among themselves and continuously congesting the system with the validation of adversarial transactions. 

If $tx$ is sound, correct, authenticated, and its owner has a balance compliance to cover the fees, the validator signs $h$, and sends the signature to its owner, who then includes the validator's signature inside $\sigma$. For a transaction $tx$ to be considered as validated, PoV requires the owner to include $\signatureThreshold$ valid signatures issued by the validators in the $search\_proof$. The validated $tx$ transaction is inserted as a Skip Graph node by its owner, which makes it accessible by other participating peers of the system to be included in a block. The numerical ID of $tx$ is $h$, and the name ID of $tx$ is $\prev$, which enables any Skip Graph peer to conduct a search for name ID on the hash value of any ledger's block within the Skip Graph overlay and find all the new transactions that are pointing back to that block. 

\label{lightchain:blk_validation}
\subsubsection{Block Generation and Validation}
We call a peer that generates blocks, a block owner. Once a block owner collects at least \txnum newly generated transactions that have not been included into any validated block that has been committed to the blockchain, it casts them into a new block $blk$, and sends the block for validation. By casting transactions into $blk$ we mean including the collected transactions into $\mathcal{S}$ set as discussed in Equation \ref{lightchain:eq_hash_blk}. \txnum is an application-dependent fixed-value parameter of \textit{\lightchain} denoting the minimum number of the transactions that should be included in a block. In contrast to the transaction owners that have more flexibility on choosing their transaction's $\prev$ pointer, the block owners should always set the $\prev$ pointer of their block to the current tail of the blockchain. Similar to the transactions, in PoV we say a block $blk$ is validated if its hash value (i.e., $h$) is being signed by $\signatureThreshold$ randomly chosen PoV validators. 
To have $blk$ validated, the block owner computes the numerical ID of the $i^{th}$ validator as shown by Equation \ref{lightchain:eq_block_validator}. 
\begin{equation}
    v_{i} = H(\prev||owner||\mathcal{S}||i) \label{lightchain:eq_block_validator}
\end{equation}
Similar to the transaction validation, the block owner searches for validators in the Skip Graph overlay, and completes the structure of $blk$ by including the search proof for validators into $search\_proof$, computing the block's hash value (i.e., $h$), and including its own signature over $h$ into $\sigma$. The block owner then contacts each of the validators and asks for the validation. Consistent with the transaction validation, a block owner is only allowed to iterate over Equation \ref{lightchain:eq_block_validator} for $i \in [1, \validatorThreshold]$. As the validation result for $blk$, the block owner either receives a signature over $h$ or $\bot$ from each contacted validator. If the block owner receives $\signatureThreshold$ signatures over $h$ from its PoV validators, it is said that the block passed the PoV validation. 
On receiving a validation request for a block $blk$, each of its PoV validators checks the authenticity and consistency of $blk$ itself, as well as the authenticity and soundness of all transactions included in $\mathcal{S}$ (as discussed earlier). The authenticity evaluation of blocks is done similar to the transactions.

\begin{figure} 
\centering
\includegraphics[scale = 0.3]{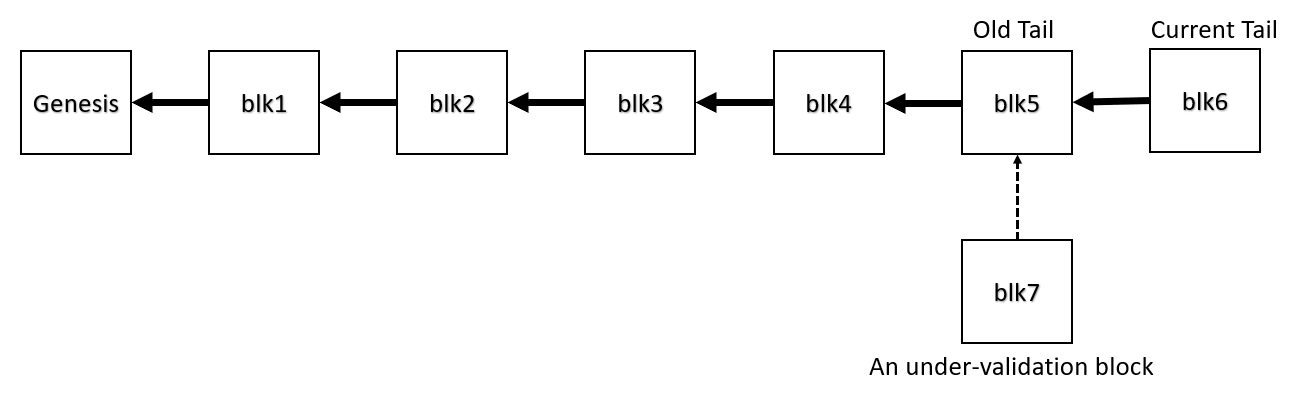}
\caption{An example of a potential fork. Validation of block $blk7$ is rejected and terminated by its validators at any state of the validation upon detection of the new block, $blk6$,  as the new tail of the blockchain.} 
\label{figure:lightchain_potential_fork}
\end{figure}

\textbf{Consistency: }A block $blk$ is said to be consistent, if its $\prev$ pointer points to the current tail of the blockchain; otherwise it is inconsistent. By the current tail of the blockchain, we mean the most recent view of the validators towards the tail of the chain. The inconsistencies among validators' views are handled by our proposed \textit{fork-free mechanism} later. However, it is likely for the current tail of the blockchain to be updated during the validation of a newly generated block. Although randomly chosen PoV validators of a block evaluate its consistency during the validation phase, nevertheless, the update on the current tail of the blockchain makes the block inconsistent during the validation procedure. Validating such an inconsistent block emerges a fork on the blockchain. To tackle this problem, once any of the randomly chosen PoV validators detects a potential fork at any step of the validation, it terminates the validation with a rejection, informing the owner. By a potential fork, we mean the situation where another block outpaces an under-validation block and becomes the new tail of the blockchain. This implies that the validators of a block need to keep their view of the blockchain's tail updated by continuously performing a search for the name ID of the hash value of the current tail (during the validation process only), which returns all the blocks and transactions that immediately succeed the tail. In this manner, upon any update on the current tail, the new tail is returned as the result of the search.
A potential fork example is illustrated by Figure \ref{figure:lightchain_potential_fork} where 
$blk7$ is undergoing the validation but its validation is terminated with rejection as soon as any of its randomly chosen PoV validators detects that another block (i.e., $blk6$) has outpaced $blk7$ in validation and became the new tail block of the blockchain.

If the block's structure is authenticated, consistent, and all the transactions in $\mathcal{S}$ are sound and authenticated, the validator signs $h$, and sends the signature to the block's owner, who includes the validator's signature inside $\sigma$. PoV considers a block $blk$ as validated if $\sigma$ field contains $\signatureThreshold$ valid signatures on $h$ value. 
After the $blk$ gets validated, its owner inserts it into the Skip Graph overlay as a node. As the incentive mechanism of \textit{\lightchain}, owner of a block receives a block generation reward once its block gets validated and committed to the blockchain. The block generation reward is a fixed-value parameter of \textit{\lightchain} that acts both as an incentive mechanism for encouraging the peers to participate progressively in generating blocks, as well as a mean for wealth creation. In this paper, we assume that the generation reward for a block is larger than its validation and routing fees. This is done to enable peers to participate in the block generation immediately after they join the system.

\begin{figure} 
\centering
\includegraphics[scale = 0.3]{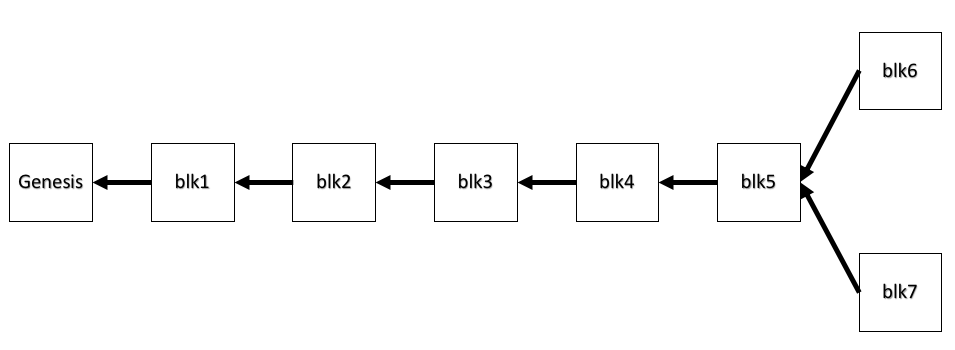}
\caption{Using our \textit{fork-free} mechanism, whichever of the simultaneously validated $blk6$ or $blk7$ has the lowest hash value wins the fork, is followed by every participating peer. The knocked-out block owners remove their blocks from the Skip Graph overlay, update their transactions set, and restart the validation procedure.
} 
\label{figure:lightchain_simultanous_mining}
\end{figure}

\textbf{Fork-free mechanism:}
To resolve the forks caused by the simultaneously validated blocks, \textit{\lightchain} governs a fork-free mechanism, which is a deterministic approach that instructs all the peers to solely follow the block with the lowest hash value upon a fork. For example, in the snapshot of Figure \ref{figure:lightchain_simultanous_mining} in the fork that is caused by simultaneous validations of the blocks $blk6$ and $blk7$ by disjoint set of PoV validators, whichever of $blk6$ or $blk7$ that has the lowest hash value is presumed as the one committed to the blockchain, and is followed by all the peers of the system. Upon a fork, we call the block with the lowest hash value as the winner block, and the other participating blocks of the fork as the knocked-out ones. The knocked-out block owners remove their block from the Skip Graph overlay, update their set of transactions by dropping the transactions that are included in the winner block, adding the new transactions to reach the \txnum threshold, and restart the validation procedure. The knocked-out block owners neither gain any block generation reward nor lose any balance because of the fees, as these fees and rewards are not effective unless the block is successfully committed to the blockchain, i.e., the block passes the PoV validation, wins the possible forks, and is appended to the current tail of the blockchain. In order to ensure that a newly appended validated block $blk$ to the ledger does not undergo any further fork rivalry, and is considered committed, effective, and finalized, \textit{\lightchain} waits for only one further block to be appended subsequently to $blk$. In this way, all the forks at the depth of the $blk$ are considered as potential forks, and are rejected by the consistency checking mechanism of PoV. Once a block $blk$ is committed to the blockchain, the contributions and fees of transactions in $\mathcal{S}$, as well as the fees and rewards associated with $blk$ itself become effective.

\subsection{\textbf{Storage Layer: Replication for better efficiency and availability}}
In \textit{\lightchain}, each transaction or block is stored in the local storage of its associated owner, and presented as a Skip Graph node, which makes it efficiently searchable by all the participating peers in the system. Hence, peers do not need to store or download the entire ledger. Rather, they access the transactions and blocks in an on-demand manner, i.e., a peer searches for a transaction or block upon a need and retrieves it efficiently from the overlay.
For further efficiency, a transaction owner should remove its transaction from the overlay once it is included in a committed block, to be discarded from the list of transactions that are waiting to be placed into the blocks. We assume the peers in \textit{\lightchain} are subject to churn, i.e., volatile between online and offline states \cite{stutzbach2006understanding}. To provide availability of the transactions and blocks under churn, all the randomly chosen PoV validators of a transaction or block also act as its corresponding replicas by storing a copy of it in their local storage, representing it as a node in the overlay, and being responsive to the other peers' queries over it. In Section \ref{lightchain:sec_results}, we show that the \textit{Signatures Threshold} parameter of \textit{\lightchain} (i.e., $\signatureThreshold$) is chosen in a way that it results in at least one available replica for each transaction and block under churn, in expectation.

\subsection{\textbf{View Layer: Randomized Bootstrapping, trusted, consistent, and efficient view synchronization}}
\label{lightchain:subsection_bootstrap}
View of a peer in \textit{\lightchain} is a table of $(numID,\, lastblk,\, state,\, balance)$ tuples. Each view table entry represents a single peer of the system with the numerical ID of $numID$, the current state of assets that is determined by the $state$, and the remaining balance of  $balance$. The $lastblk$ represents the hash value of the last block on the blockchain that contains the most recent transaction of that peer. We define the \textit{view introducers} of a new peer as the set of randomly chosen peers that share their view of the blockchain with the newly joined peer. Upon joining the overlay, a new peer computes the numerical IDs of its view introducers based on Equation \ref{lightchain:eq_view_introducers}, where $new\_peer.numID$ is the numerical ID of the new peer and $view\_intro_{i}$ is the numerical ID of the $i^{th}$ view introducer of it. We employ the hash function as a random oracle to obtain uniformly random view introducers' numerical IDs. 
\begin{equation}
   view\_intro_{i} = H(new\_peer.numID||i)    
\label{lightchain:eq_view_introducers}
\end{equation}
The new peer then conducts a search for numerical ID of $view\_intro_{i}$ within the overlay, contacts the peer in the search result, and obtains its view of the blockchain. The new peer continues in this manner by iterating over $i$ until it obtains $\signatureThreshold$ consistent views. As we show later, we determine  $\signatureThreshold$ and $\validatorThreshold$ in such a way that a new peer obtains $\signatureThreshold$ consistent views of the honest peers by iterating $i$ over $[1, \validatorThreshold]$. 

\subsection{Direct retrieval of the latest state} 
\label{lightchain:subsection_direct_access}
Each transaction that appears on a committed block to the ledger contains the latest update on the transaction owner's state of assets. \textit{\lightchain}'s approach on representing each block by a Skip Graph node makes the blocks addressable, searchable, and efficiently retrievable within the network. Tracking the updates on the entire view of other peers' assets, hence, requires a peer to keep its local view updated with the new blocks, which is a plausible assumption in the majority of the existing solutions. However, in addition to sequentially seeking the new blocks and updating view accordingly, \textit{\lightchain} enables each peer to directly retrieve the latest assets' state of another peer of interest without the need to keep track of the new blocks on the ledger. This is done by the additional representation of each block with multiple Skip Graph nodes i.e., one per each transaction included in the block. As each of these additional Skip Graph nodes represents one of the transactions of the same block, we call them the associated \textit{transaction pointers} of that block. In this approach, each transaction $tx$ that is included into a committed block $blk$ is represented by a transaction pointer node (i.e., $pointer$). The name ID and numerical ID of the transaction pointer node are set as $pointer.nameID = tx.owner$ and $pointer.numID = blk.h$, respectively. Setting the numerical ID of a transaction pointer to its associated block's hash value is for the sake of security, and to provide a tie between each pointer and the block it points to. The transaction pointer nodes associated with each block are inserted by the block's owner and replicated on the block's PoV validators. In this manner, a peer that is solely interested in knowing the latest state of another peer's assets, for example, $tx.owner$, performs a search for a transaction pointer with the name ID of $tx.owner$ as the search target within the Skip Graph overlay. The search is answered by either the owner of $blk$ or one of its PoV validators that all keep a copy of $blk$ (i.e., the block that contains the latest update on the assets' state of $tx.owner$). 
To keep track of the latest updates over the assets, both the owner and validators of a block should take down each of its associated transaction pointers once an update on the corresponding assets appears on a newer committed block to the ledger. Taking down a pointer node from the overlay is simply done by performing the Skip Graph node deletion operation \cite{aspnes2007skip} by the owner and each of the validators in a fully decentralized manner. This is for the sake of better efficiency of the search, and to make sure that the transaction pointers always point to the most recent states. Not dropping the pointers after a new update is counted as misbehavior, which we address it by the misbehavior detection strategy of \textit{\lightchain}. To address the network asynchrony, however, the block owner and PoV validators are allowed to take down the pointers within at most a certain number committed blocks after a new transaction on the associated set of assets happens. This allows them to have enough time to discover the new updates without being subject to misbehavior. The length of the block interval (i.e., number of blocks between two transaction pointers over the same set of assets) is a constant protocol parameter that is application dependent.

\subsection{\textbf{Motivating honest behavior and misbehavior detection:}}
\label{lightchain:subsection_auditing}
The block generation reward and the routing and validation fees constitute the incentive mechanism of the \textit{\lightchain} for the peers to retain their honest behavior i.e., following the \textit{\lightchain}'s architecture and protocol as described in this section. In this paper, we assume that the block generation reward is greater than the routing and validation fees. We establish this assumption to motivate any peer to retain honest behavior from the time it joins the system by enabling it to participate in block generation and gain the block generation reward. The counterpart of honest behavior is the \textit{misbehavior}, which we define it as any sort of deviation from the described \textit{\lightchain}'s protocol and architecture. As detailed earlier, for the transactions and blocks that are gone through the consensus layer, we consider the randomly chosen PoV validators to check the submitted transaction or block against the misbehavior. As we analyze in Section \ref{lightchain:sec_results}, we choose the system's security parameter (i.e., $\lambda$) as well as PoV operational parameters (i.e., $\validatorThreshold$ and $\signatureThreshold$) in a way that an adversarial peer cannot convince the PoV validators on a misbehavior unless with a negligible probability $\epsilon(\lambda)$ (e.g., $\epsilon(\lambda) = 2^{-\lambda}$). In addition to the countermeasures established by PoV, we also introduce \textit{misbehavior detection} as an extra level of adversarial countermeasure, especially for the adversarial actions that are not gone through the PoV e.g., direct submission of an invalid block to the ledger. In our proposed misbehavior detection, 
each peer of \textit{\lightchain} acts as an auditor for other peers' behavior and gains a \textit{misbehavior audition reward} by reporting their misbehavior. As an auditor, any peer should be able to evaluate a block or transaction in the same way that its PoV validators do during the validation. Any deviation from \textit{\lightchain}'s protocol that fails the auditor's evaluation is considered as misbehavior, e.g., the invalid signature of validators, lack of $\signatureThreshold$ signatures on the hash value, and validating an invalid block or transaction. We specified the first two cases earlier in this section. The last case (i.e., validating an invalid block or transaction) happens when an adversarial transaction or block owner finds $\signatureThreshold$ randomly chosen malicious PoV validators who deviate from the validation protocol and sign an invalid block or transaction, e.g., a double-spending transaction. Although as we stated earlier, we determine PoV operational parameters in a way that such an attack cannot happen unless with a negligible probability of at most $\epsilon(\lambda)$, nevertheless, the misbehavior detection of \textit{\lightchain} provides an extra level of security to ensure that even if such an attack happens, the invalid transaction or block does not persist on the ledger.

Upon a misbehavior detection, the auditor generates a transaction with the evidence of the misbehavior in the contribution field. The transaction then goes through the same PoV validation process as described earlier, except that the validators verify the correctness of the transaction as the correctness of the reported evidence. Once the transaction is validated and placed into a committed block to the blockchain, the guilty peer is penalized by the misbehavior penalty fee, routing fee, and validation fee that it is made to pay to the owner (i.e., auditor), routers, and validators of the transaction, respectively. Misbehavior fee is another constant parameter of \textit{\lightchain} that is application dependent. Once misbehavior is recorded for a peer on a committed block, its identifier is blacklisted. The blacklisted peers are isolated by honest peers i.e., any incoming message from the blacklisted peers is discarded by honest peers. This eventually results in the blacklisted peers being excluded from the overlay, which causes the blacklisted peers to never being selected as a validator as they no longer are a part of the overlay from the honest peers' point of view. A blacklisted peer appearing in an authenticated search proof implies a malicious router peer on the search path that is caught and blacklisted accordingly.

The detailed pseudo-code descriptions of \textit{\lightchain}'s algorithms are presented in the Appendix.

\section{Security Analysis}
\label{lightchain:sec_security}
In this section, we formalize the security analysis of \textit{\lightchain} at the protocol-level, and derive the necessary mathematical conditions on determining its operational system parameters: the \textit{Signatures Threshold} $\signatureThreshold$, and the \textit{Validators Threshold} $\validatorThreshold$. We evaluate the security of \textit{\lightchain} from the lens of classical information security standard of \textit{C.I.A}, which stands for the \textit{Confidentiality}, \textit{Integrity}, and \textit{Availability} \cite{goodrich2011introduction}. Among the aforementioned features, we scope out confidentiality as we aim at \textit{\lightchain} operations to be transparent and verifiable by everyone publicly at the plaintext level. Hence, we leave confidentiality as an application-dependent feature, e.g., confidentiality-preserving password storage applications. We scope the security of \textit{\lightchain} in providing the Integrity and Availability features. In the definition of integrity and availability, we mainly consider the Consensus, Storage, and View layers of \textit{\lightchain}, and scope out the Network layer availability and integrity, as both features are reduced to the availability, and integrity of the secure routing protocol that \textit{\lightchain} is operating on (e.g., \cite{boshrooyeh2017guard}).

\textbf{Adversarial and Honest Behavior:} We assume all participating peers in \textit{\lightchain} system as probabilistic Turing machines that run in a polynomial time in the security parameter of the system (i.e., $\lambda$). We define honest behavior as the one follows \textit{\lightchain} protocols as specified in Section \ref{lightchain:sec_solution}, and maintains its availability and accessibility in the system. On the contrary, we define adversarial behavior as the one that deliberately deviates from the \textit{\lightchain} protocol collectively at arbitrary points. Similar to BFT-based approaches \cite{pease1980reaching}, we assume the existence of a Sybil adversarial party \cite{douceur2002sybil} that adaptively takes control over a fraction of at most $\adv$ of (corrupted) peers in the system. The adversarial party aims at compromising the integrity or availability of \textit{\lightchain} by orchestrating attacks through its under-controlled set of corrupted peers. In our analysis, to presume the worst-case scenario, we assume no churn for the corrupted peers under the control of the adversary, i.e., corrupted peers never leave the system. For the honest peers, on the other hand, we model the churn with a uniform failure probability of $\ufp$, under crash-recover model \cite{tanenbaum2007distributed}, i.e., an honest node fails with a probability of $\ufp$ and comes back online after a while. Based on these assumptions, the expected number of online peers at any time in the system is denoted by $n_{o}$ and determined by the Equation \ref{lightchain:eq_online_peers}. As is represented later in Section \ref{lightchain:sec_results}, we verified these assumptions against practical churn traces in our simulation. 

\begin{equation}
    n_{o} = \systemCapacity (\adv + (1 - f)(1 - \ufp))
    \label{lightchain:eq_online_peers}
\end{equation}

\subsection{Integrity}
Integrity is generalized as the property of the information not being altered in an unauthorized fashion \cite{goodrich2011introduction}. In the context of \textit{\lightchain} we
define the \textit{integrity} as the property that the views of the peers towards the blockchain are not being changed, except by committing a new block of validated transactions to the current tail of the blockchain solely by the designated PoV validators of that block. We further break down the integrity feature into the integrity of Consensus and View layers, and assume the integrity of \textit{\lightchain} is preserved as long as the integrity of both its View and Consensus layers preserved. 

\textbf{Consensus Layer Integrity:} The integrity of the Consensus layer is preserved as long as the integrity of all committed transactions and blocks to the ledger is preserved. We define the integrity of a committed transaction as the property that preserves its Soundness, Correctness, Authenticity, and Balance Compliance features altogether over time as defined in Section \ref{lightchain:tx_validation}. A committed transaction violates integrity if it violates any of the aforesaid features. Similarly, we define the integrity of a committed block as the property that preserves its Authenticity and Consistency as well as the integrity of all transactions it contains as defined in Section \ref{lightchain:blk_validation}. A committed block violates the integrity feature if either it does not have Authenticity or Consistency, or any of the transactions it contains violates the integrity. According to these definitions, all adversarial attacks aiming at unauthorized tempering of the ledger in \textit{\lightchain} ends up breaking its integrity, e.g, double-spending (i.e., transaction soundness violation), forking history (i.e., block consistency violation), modifying order of blocks or their content (i.e., block consistency violation) \cite{li2020survey}. At the Consensus layer, any transaction or block made available on underlying Skip Graph of \textit{\lightchain} should pass the PoV validation successfully, and otherwise is discarded and considered misbehavior subjected to blacklisting (see Section \ref{lightchain:subsection_auditing}). No honest PoV validator approves an integrity-violating transaction or block. Also, nodes in \textit{\lightchain} are utilizing an existentially unforgeable digital signature scheme under adaptive chosen-message attack \cite{katz2014introduction}, which makes it computationally infeasible for any probabilistic polynomial time adversary to forge signatures on behalf of assigned PoV validators. Hence the only way for an adversary to break the Consensus layer integrity of \textit{\lightchain} is to find $\signatureThreshold$ \textit{corrupted} PoV validators out of $\validatorThreshold$ trials based on Equations \ref{lightchain:eq_validator} and \ref{lightchain:eq_block_validator} to vote in the favor of its \textit{corrupted} transaction or block, respectively. By a corrupted transaction or block, we mean the one that violates the (Consensus layer) integrity of \textit{\lightchain}.

\textbf{View Layer Integrity:} The integrity of the View layer is preserved as long as the view of every honest peer at the end of bootstrapping to \textit{\lightchain} is identical to the view that is constructed by applying all committed blocks to the ledger from the Genesis to the tail of the ledger at the start of bootstrapping. According to this definition, the View layer integrity is violated if the view of an honest peer at the end of bootstrapping to the system gets any different than the view constructed by replaying all blocks from Genesis to the current tail at the time the peer joined. By the bootstrapping phase, we mean a successful execution of the Randomized Bootstrapping protocol of \textit{\lightchain} as defined in Section \ref{lightchain:subsection_bootstrap}. The only way for an adversary to break the View layer integrity is to impose $\signatureThreshold$ \textit{corrupted} view introducers to an honest node out of its $\validatorThreshold$ trials based on Equations \ref{lightchain:eq_view_introducers} during its Randomized Bootstrapping protocol, which results in the honest node being bootstrapped with a corrupted view. By a corrupted view, we mean the one that violates the View layer integrity of \textit{\lightchain}.

\begin{theorem}
\label{lightchain:thm_integrity}
Given a \textit{\lightchain} system of $\systemCapacity$ peers with a partially synchronous network, an authenticated routing mechanism (as defined in Section \ref{lightchain:sec_preliminaries}), where honest nodes follow a uniform failure probability of $\ufp$ under the crash-recover model, and considering the negligible function $\epsilon(\lambda)$, there is a configuration of \textit{\lightchain}'s parameters that yields the success probability of an adversary controlling a fraction $\adv$ of all peers on breaking the integrity of \textit{\lightchain} bound by $\epsilon(\lambda)$.
\end{theorem}
\textit{Proof.} 
Let $V_{\validatorThreshold}$ be a random variable denoting the number of corrupted peers within at most $\validatorThreshold$ trials during PoV validation or Randomized Bootstrapping protocols. To provide both View and Consensus layer integrity for \textit{\lightchain} we aim at achieving $Pr(V_{\validatorThreshold} \geq \signatureThreshold) \leq \epsilon(\lambda)$, i.e., the probability of achieving $\signatureThreshold$ corrupted peers within at most $\validatorThreshold$ trails is bound by the negligible function $\epsilon(\lambda)$. 
Equation \ref{lightchain:eq_prob_malicious_validators} represents the evaluation of the cumulative distribution function of $V_{\validatorThreshold}$ at $\signatureThreshold$. For the large values of $\systemCapacity$, Equation \ref{lightchain:eq_prob_malicious_validators} is approximated by Equation \ref{lightchain:eq_p_psi}, which denotes a Hypergeometric distribution \cite{harkness1965properties}, where $\psi$ is the standard normal distribution function \cite{bertsekas2002introduction}. Equation \ref{lightchain:eq_t_m} represents a lower bound on the value of $\signatureThreshold$ that results in $Pr(V_{\validatorThreshold} \geq \signatureThreshold) \leq \epsilon(\lambda) $. The lower bound obtained from this equation is independent of the system capacity (i.e., $\systemCapacity$) as well as their churn (i.e., $\ufp$). Having Equation \ref{lightchain:eq_t_m} and given the assumption that $\validatorThreshold \geq \signatureThreshold$, a lower bound value for $\validatorThreshold$ is obtained as shown by Equation \ref{lightchain:alpha_lower_bound}. Hence, given adversarial fraction of system $\adv$, and the security parameter $\lambda$, Equations \ref{lightchain:eq_t_m} and \ref{lightchain:alpha_lower_bound} determine the range of operational parameters of \textit{\lightchain} for which both its Consensus and View layer integrity are held with a very high probability in the security parameter of the system (i.e., probability of $1-\epsilon(\lambda)$). In other words, the minimum value of $\validatorThreshold$ for preserving the integrity of system is obtained from Equation \ref{lightchain:alpha_lower_bound}, and for a certain value of $\validatorThreshold$, choosing $\signatureThreshold$ based on Equation \ref{lightchain:eq_t_m} results in a probability of at most $\epsilon(\lambda)$ to have $\signatureThreshold$ corrupted peers within at most $\validatorThreshold$ trials to use them as PoV validators of a corrupted transaction or block and break the Consensus layer integrity of \textit{\lightchain}, or impose them as the view introducers to an honest peer and break the View layer integrity of \textit{\lightchain}. This concludes the proof. $\blacksquare$

\begin{align}
    Pr(V_{\alpha} < t) &= \frac{\sum_{i=0}^{t - 1} \binom{\systemCapacity\adv}{i} \binom{\systemCapacity(1-\adv)(1-\ufp)}{\validatorThreshold-i}}{\binom{n_{o}}{\validatorThreshold}}
    \label{lightchain:eq_prob_malicious_validators}\\
    & \approx \psi(\frac{t - \validatorThreshold \adv - 1}{\sqrt{\validatorThreshold \adv(1-\adv)}})
    \label{lightchain:eq_p_psi}\\
     t \geq (\sqrt{\validatorThreshold \adv(1-\adv)} &\times \psi^{-1}(1- \epsilon(\lambda))) + \validatorThreshold \adv + 1
     \label{lightchain:eq_t_m}\\
     \validatorThreshold &\geq \frac{(\sqrt{ \adv} + \sqrt{\adv \times (\psi^{-1}(1- \epsilon(\lambda)))^{2} + 4}))^{2}}{4(1-\adv)}
     \label{lightchain:alpha_lower_bound}
\end{align}

\subsection{Availability}
We decouple the availability feature of \textit{\lightchain} into the \textit{data availability} and the \textit{service availability}. 

\subsubsection{Data Availability}
We define the data availability as the property that all validated transactions and blocks being accessible by all participating peers within a strict time-bound $\zeta$, i.e., the time it takes for a requested validated transaction or block to be made available at the requester's disposal is bound by $\zeta$. The data availability feature is violated when there is no such $\zeta$ time-bound available, i.e., a transaction or block is permanently deleted from the system. Since in \textit{\lightchain} PoV validators of a transaction or block serve as its replicas, violating the data availability corresponds to an adversary managing all validators of a transaction or block to be selected among its set of corrupted peers, so that it takes full control over the availability of that transaction or block, and makes it unavailable. Otherwise, as long as a transaction or block has at least an honest PoV validator, its data availability is preserved. Note that following the crash-recover assumption of nodes in \textit{\lightchain}, we do not consider churn of honest replicas compromising data availability since we assume a departed or failed replica gets back to the system after some certain time. Hence choosing $\zeta$ long enough to account for the unavailability period of the honest replicas preserves the data availability of \textit{\lightchain}. 

\begin{theorem}
\label{lightchain:thm_data_availability}
Given a \textit{\lightchain} system of $\systemCapacity$ peers, a partially synchronous network, an authenticated routing mechanism (as defined in Section \ref{lightchain:sec_preliminaries}), where honest nodes follow a uniform failure probability of $\ufp$ under the crash-recover model and considering the negligible function $\epsilon(\lambda)$, as well as a probabilistic polynomial-time adversary that controls a fraction $\adv$ of peers, there is a configuration of \textit{\lightchain}'s parameters such that the data availability is preserved with a very high probability in the security parameter of the system, i.e., the probability of at least $1- \epsilon(\lambda)$.
\end{theorem}
\textit{Proof.} We prove the theorem in two steps. In the first step, we show that there is a configuration of a system that makes at least one honest replica (i.e., validator) for every validated data object (i.e., transaction or block) with a very high probability in the security parameter of the system. In the second step, we prove that even with the minimum of one honest replica for a validated data object, the data availability is preserved in the sense that the validated data object never goes indefinitely unavailable, i.e., there exists a strict time-bound $\zeta$ for which requesting the validated data object is answered by its replica within at most $\zeta$ time units with a very high probability in the security parameter of the system. 

We define $\Xi$ as the event in which at least one replica for a data object is chosen from the honest peers. Following Theorem \ref{lightchain:thm_integrity}, choosing  $\signatureThreshold$ and $\validatorThreshold$  based on Equations \ref{lightchain:eq_t_m} and \ref{lightchain:alpha_lower_bound}, respectively, yields in event $\Xi$ happening with a high probability in the security parameter of the system, i.e., $Pr[\Xi] \geq 1 - \epsilon(\lambda)$. 
Assume that all other (corrupted) validators permanently drop the validated data object from their storage right after they validate it, hence aiming at violating the data availability of the system. This leaves only a single copy of that data object in the system, which is maintained by its sole honest validator. Following the assumption of honest nodes showing a churn behavior with a crash-recover probability of $\ufp$, the sole honest replica of the validated data object may either be online or offline at the time another node requests the validated data object it holds. If the node is offline, the probability that it comes online after some finite $\zeta_{1}$ time units follows a geometric distribution with the success probability of $1-\ufp$. We denote the event the offline honest replica comes online after $\zeta_{1}$ time units by $\Xi'_{\zeta_{1}}$. As shown by Equation \ref{lightchain:eq_zeta_offline} we aim at event $\Xi'_{\zeta_{1}}$ happen with a very high probability in the security parameter of the system, which yields in the upper bound limit on $\zeta_{1}$ as shown by Equation \ref{lightchain:eq_zeta_offline_bound}. This equation implies that any honest offline replica comes online within at most $\zeta_{1}$ time units with a very high probability in the security parameter of the system. Once the replica is online it answers any request on the data objects it holds within a limited time bound of $\zeta_{2}$ as shown by Equation \ref{lightchain:eq_zeta_online}, where $\Delta$ is the maximum message propagation time between two nodes in the partially synchronous \textit{\lightchain} system and $\delta(\lambda)$ is the local processing computation time overhead of replica on answering the request. Accordingly, we derive the value of $\zeta$ as $\zeta_{1} + \zeta_{2}$, which implies that every validated data object in our proposed \textit{\lightchain} is accessible within a finite time, i.e., no validated data object goes unavailable indefinitely. This concludes the data objects' availability proof. $\blacksquare$
\begin{align}
    Pr[\Xi'_{\zeta_{1}}] = \ufp^{\zeta_{1} - 1}\times (1-\ufp) &\geq 1 - \epsilon(\lambda) \label{lightchain:eq_zeta_offline}\\
    \zeta_{1} & \leq \log_\ufp{\frac{1 - \epsilon(\lambda)}{1-\ufp}} \label{lightchain:eq_zeta_offline_bound}\\
    \zeta_{2} &\leq 2 \times \Delta + \delta(\lambda) \label{lightchain:eq_zeta_online}
\end{align}

\subsubsection{Service Availability}
We define the service availability as the availability of the Consensus and View layer protocols of \textit{\lightchain} on validating the transactions and blocks, as well as providing the Randomized Bootstrapping services, respectively. 

\textbf{Consensus Layer Service Availability:} Having the Consensus layer service availability, each honest peer in \textit{\lightchain} should be able to get its honest inquiry validated successfully within one round of trials in expectation. A round of trials is limited to trying at most $\validatorThreshold$ designated validators by the PoV protocol. By an honest inquiry, we mean any transaction or block that is evaluated as valid within any PoV round of trials that contains $\signatureThreshold$ honest validators. The Consensus layer service availability is violated when the adversary succeeds to \textit{indefinitely} hold back an honest node from validating its honest inquiry. This happens when the adversary succeeds in indefinitely imposing more than $\validatorThreshold - \signatureThreshold$ corrupted PoV validators assigned to an honest node's validation inquiry. The corrupted assigned validators keep rejecting honest inquiry, hence, indefinitely hindering back the honest node from progressing in the system in terms of generating transactions and blocks. 
Note that we emphasize the success of adversary on indefinitely-many rounds of trial, as being failed to find enough validators within a single round, the honest node may try tweaking a nonce in the script of its transaction or block to find another set of PoV validators. As long as the honest node can succeed in finding enough honest validators to approve its correct inquiry in expectation within a single round of trials, the Consensus layer service availability is preserved.

\textbf{View Layer Service Availability:} Having the View layer service availability, each participating peer should be able to bootstrap itself with the Randomized Bootstrapping protocol within a single round of trials in expectation (similar to the Consensus layer availability). A round of trials is limited to trying at most $\validatorThreshold$ view introducers by the Randomized Bootstrapping protocol. Likewise, violating the View layer service availability happens when the adversary succeeds in indefinitely imposing more than $\validatorThreshold - \signatureThreshold$ corrupted view introducers to an honest node's Randomized Bootstrapping inquiry. The corrupted assigned view introducers keep rejecting the bootstrapping request of the honest node, hence, indefinitely hindering back the honest node from bootstrapping to the system. Similar to the Consensus layer service availability, we emphasize the success of adversary on indefinitely-many rounds of trials, as being failed to find enough view introducers within a single round, the honest node may try tweaking a nonce in the script of its request to find another set of view introducers. As long as the honest node is able to succeed in finding enough view introducers and bootstrap to the system within a single round of trials in expectation, the View layer service availability is preserved.

Note that in \textit{\lightchain} having the Consensus and View layers service availability are the deterrence countermeasures against an adversary hindering at least a single honest node on progressing with the consensus or bootstrapping to the system, respectively. This also corresponds to the liveness feature in the distributed systems paradigm \cite{tanenbaum2007distributed}. However, the service availability does not account for an adversary trying to slow down the progress of the system by \textit{occasionally} succeeding in blocking a node's progress in consensus or bootstrapping within a limited number of rounds.

\begin{theorem}
\label{lightchain:thm_service_availability}
Given a \textit{\lightchain} system of $\systemCapacity$ peers, with a partially synchronous network, an authenticated routing mechanism (as defined in Section \ref{lightchain:sec_preliminaries}), where honest nodes follow a uniform failure probability of $\ufp$ under the crash-recover model, as well as a probabilistic polynomial-time adversary that controls a fraction $\adv$ of peers, there is a configuration of \textit{\lightchain}'s parameters such that the service availability is always preserved.
\end{theorem}

\textit{Proof.} Based on the definitions of service availability for the Consensus and View layers, it is inferred that the service availability feature of \textit{\lightchain} is generalized as the property that every honest peer should be able to find a minimum of $\signatureThreshold$ honest peers within a single round of trials in expectation, where each round involves sampling $\validatorThreshold$ peers uniformly following the Consensus or View layer protocols of \textit{\lightchain}. To prove the theorem, we determine the value of $\signatureThreshold$ in such a way that finding $\signatureThreshold$ \textit{honest} peers is achievable in expectation within a single round $\validatorThreshold$ trials. As shown by Equation \ref{lightchain:eq_prob_honest_validators}, we denote by $p$ the Bernoulli probability of choosing an honest peer uniformly at random. The uniform distribution of $p$ follows from the random oracle model of hash function employed in the PoV and Randomized Bootstrapping protocols. Given a certain value of $\validatorThreshold$, we denote variable $H_\validatorThreshold$ as the random variable corresponding to the number of honest peers obtained within a round of $\validatorThreshold$ trials. Following Theorem \ref{lightchain:thm_integrity}, $H_\validatorThreshold$ follows a Hypergeometric distribution with an expected value as shown by Equation \ref{lightchain:eq_expected_honest_validators}, which denotes the expected number of honest peers within a single round of trials. We aim at choosing $\signatureThreshold$ in such a way that it enables an honest peer to find $\signatureThreshold$ honest peers within at most $\validatorThreshold$ trials in expectation. This results in an upper bound value of $\signatureThreshold$ as shown by Equation \ref{lightchain:eq_expected_honest_validators_upper_bound}. For a certain value of $\validatorThreshold$, choosing $\signatureThreshold$ based on Equation \ref{lightchain:eq_expected_honest_validators_upper_bound} results in having a minimum of $\signatureThreshold$ honest peers in a single round of trials in expectation. This concludes the service availability proof. $\blacksquare$

\begin{align}
    p&= \frac{\systemCapacity \times (1-\adv) \times (1- \ufp)}{n_{o}}
    \label{lightchain:eq_prob_honest_validators} \\
    E[H_\validatorThreshold] &= \validatorThreshold \times p
    \label{lightchain:eq_expected_honest_validators}\\
    \signatureThreshold &\leq \validatorThreshold \times p
    \label{lightchain:eq_expected_honest_validators_upper_bound}
\end{align}

\section{Performance Results}
\label{lightchain:sec_results}
\subsection{Simulation Results}
\textbf{Setup: } To simulate and evaluate the integrity as well as data and service availability of \textit{\lightchain}, we implemented \textit{\lightchain} over the Skip Graph simulator SkipSim \cite{hassanzadeh2020skipsim}. 
To model the dynamic online and offline behavior of peers (i.e., churn) in a realistic manner, we follow the extracted churn model of the Bitcoin system \cite{imtiaz2019churn}, where each peer has the periodic states of online and offline with the expected duration of $10.6$ and $2.8$ hours, respectively. Consistent with the reported statistics of the Bitcoin provided in \cite{bitnodes, blockchaincharts}, in our \textit{\lightchain} implementation on SkipSim, while a peer is online it generates a single transaction per hour. We simulated \textit{\lightchain} for $100$ randomly generated Skip Graph overlay topologies, each with the system capacity of $\systemCapacity=10,000$ peers. By randomly generated topologies we mean the randomized set of peers' identifiers as well as the overlay connectivity. Each topology was simulated for $48$ hours, and for the sake of simulation, we set \txnum = 1. 

\begin{figure*}
\centering
\begin{minipage}[c]{.5\textwidth}
\centering
    \scalebox{0.6}{\begin{tikzpicture}
\begin{axis}[
    ylabel={Adversarial Success Probability},
    xlabel={Signatures Threshold ($\signatureThreshold$) },
    xmin=0, xmax=40,
    ymin=0, ymax=1,
    xtick={0, 5, 10, 15, 20, 25, 30, 35, 40},
    ytick={0, 0.1, 0.2, 0.3, 0.4, 0.5, 0.6, 0.7, 0.8, 0.9, 1},
    legend style={
    }, 
    ymajorgrids=true,
    grid style=dashed,
]

    \addplot
    coordinates 
    {
        (1,  .81)
        (2, .47)
        (3, .19)
        (4, .05)
        (5, .01)
        (6, 10^-3)
        (7, 10^-4)
        (8, 0)
        (9, 0)
        (10,0)
    };
    \addlegendentry{$\adv = 0.16, \validatorThreshold = 10$}
    
    \addplot
    coordinates 
    {
        (5,  .86)
        (10, .11)
        (12, .017)
        (15, 10^-4)
        (18, 0)
        (19, 0)
        (20, 0)
    };
    \addlegendentry{$\adv = 0.33, \validatorThreshold = 20$}
    
    \addplot
    coordinates 
    {
        (15, .95)
        (20, .53)
        (25, .06)
        (30, 10^-5)
        (35, 0)
        (36, 0)
        (37, 0)
        (38, 0)
        (39, 0)
    };
    \addlegendentry{$\adv = 0.51, \validatorThreshold = 39$}

\end{axis}
\end{tikzpicture}}
    \caption*{(a) Integrity}
\end{minipage}
\begin{minipage}[c]{.5\textwidth}
\centering
    \scalebox{0.6}{\begin{tikzpicture}
\begin{axis}[
    ylabel={Adversarial Success Probability},
    xlabel={Signatures Threshold ($\signatureThreshold$) },
    xmin=0, xmax=100,
    ymin=0, ymax=1,
    xtick={0, 10, 20, 30, 40, 50, 60, 70, 80, 90, 100},
    ytick={0, 0.1, 0.2, 0.3, 0.4, 0.5, 0.6, 0.7, 0.8, 0.9, 1},
    legend style={
    }, 
    ymajorgrids=true,
    grid style=dashed,
]

    \addplot
    coordinates 
    {
        (1, .81)
        (4, .17)
        (8,  10^-4)
        (10, 10^-6)
        (11, 0)
        (12, 0)
        (14, 0)
    };
    \addlegendentry{$\adv = 0.16, \validatorThreshold = 14$}
    
    \addplot
    coordinates 
    {
        (30, .9)
        (50, .3)
        (60, 10^-2)
        (70, 10^-5)
        (80, 0)
        (90, 0)
        (100, 0)
    };
    \addlegendentry{$\adv = 0.33, \validatorThreshold = 144$}

\end{axis}
\end{tikzpicture}}
    \caption*{(b) Integrity and Service Availability}
\end{minipage}
\begin{minipage}[c]{.5\textwidth}
\centering
    \scalebox{0.6}{\begin{tikzpicture}
\begin{axis}[
    ylabel={Average Blocks Availability},
    xlabel={Signatures Threshold ($\signatureThreshold$) },
    xmin=0, xmax=100,
    ymin=0, ymax=85,
    xtick={0, 10, 20, 30, 40, 50, 60, 70, 80, 90, 100},
    ytick={0, 10, 20, 30, 40, 50, 60, 70, 80, 85},
    legend style={
    legend pos=north west, legend cell align = left
    }, 
    ymajorgrids=true,
    grid style=dashed,
]
    \addplot
    coordinates 
    {
        (1,  1.6)
        (4,  4.3)
        (8,  7.5)
        (10, 9.4)
        (11, 10.4)
        (12, 11.1)
    };
    \addlegendentry{$\adv = 0.16, \validatorThreshold = 14$}
    
    \addplot
    coordinates 
    {
        (50,  43)
        (60,  51)
        (70,  60)
        (80,  69)
        (90,  77)
        (100, 85)
    };
    \addlegendentry{$\adv = 0.33, \validatorThreshold = 144$}

\end{axis}
\end{tikzpicture}}
    \caption*{(c) Data Availability}
\end{minipage}
\caption{Performance evaluation of \textit{\lightchain} concerning its integrity, and service and data availability. The performance is reported as averages over $100$ \textit{\lightchain} systems each with $10,000$ nodes. Each system is simulated for $48$ hours under the real churn traces of Bitcoin \cite{imtiaz2019churn}.}
\label{lightchain:fig_simulation_results}
\end{figure*}

\begin{figure*}
\centering
\begin{minipage}[c]{.5\textwidth}
\centering
    \scalebox{0.6}{\begin{tikzpicture}
\begin{axis}[
    ylabel={Average Block Formation Time (s)},
    xlabel={Transactions Number},
    xmin=0, xmax=40,
    ymin=0, ymax=70,
    xtick={0, 10, 20, 40},
    ytick={0, 5, 10, 15, 20, 25, 30, 35, 40, 45, 50, 55, 60, 65, 70},
    legend style={
    legend pos=north west, legend cell align = left
    }, 
    ymajorgrids=true,
    grid style=dashed,
]
    \addplot
    coordinates 
    {
        (10,  1.8)
        (20,  3.6)
        (40,  6.3)
    };
    \addlegendentry{$\adv = 0.16, \validatorThreshold = 14$}
    
    \addplot
    coordinates 
    {
        (10,  15.58)
        (20,  32.98)
        (40,  66.86)
    };
    \addlegendentry{$\adv = 0.33, \validatorThreshold = 144$}
\end{axis}
\end{tikzpicture}}
    \caption*{(a) The block formation time versus the number of transactions in a block}
\end{minipage}
\begin{minipage}[c]{.5\textwidth}
\centering
    \scalebox{0.6}{\begin{tikzpicture}
\begin{axis}[
    ylabel={Average Block Validation Time (s)},
    xlabel={Number of Transactions},
    xmin=0, xmax=40,
    ymin=0, ymax=75,
    xtick={0, 10, 20, 40},
    ytick={0, 5, 10, 15, 20, 25, 30, 35, 40, 45, 50, 55, 60, 65, 70, 75},
    legend style={
    legend pos=north west, legend cell align = left
    }, 
    ymajorgrids=true,
    grid style=dashed,
]
    \addplot
    coordinates 
    {
        (10,  9)
        (20,  18)
        (40,  35)
    };
    \addlegendentry{$\adv = 0.16, \validatorThreshold = 14$}
    
    \addplot
    coordinates 
    {
        (10,  25)
        (20,  37)
        (40,  60)
    };
    \addlegendentry{$\adv = 0.33, \validatorThreshold = 144$}
\end{axis}
\end{tikzpicture}}
    \caption*{(b) The block validation time versus the number of transactions in a block}
\end{minipage}
\begin{minipage}[c]{.5\textwidth}
\centering
    \scalebox{0.6}{\begin{tikzpicture}
\begin{axis}[
    ylabel={Average Block Size (KB)},
    xlabel={Transactions Number},
    xmin=0, xmax=40,
    ymin=0, ymax=800,
    xtick={0, 10, 20, 40},
    ytick={0, 100, 200, 300, 400, 500, 600, 700, 800},
    legend style={
    legend pos=north west, legend cell align = left
    }, 
    ymajorgrids=true,
    grid style=dashed,
]
    \addplot
    coordinates 
    {
        (10,  31)
        (20,  59)
        (40,  115)
    };
    \addlegendentry{$\adv = 0.16, \validatorThreshold = 14$}
    
    \addplot
    coordinates 
    {
        (10,  208)
        (20,  387)
        (40,  743)
    };
    \addlegendentry{$\adv = 0.33, \validatorThreshold = 144$}
\end{axis}
\end{tikzpicture}}
    \caption*{(c) The block size versus the number of transactions in a block}
\end{minipage}
\caption{The performance of the proof-of-concept of \textit{\lightchain} deployment on the Google Cloud Platform with $1000$ nodes each generating $1000$ transactions.}
\label{lightchain:fig_poc_results}
\end{figure*}

\textbf{Integrity:} Figure \ref{lightchain:fig_simulation_results}.a shows the success probability of the adversary that controls fraction $\adv$ of corrupted peers on compromising the integrity of the system. We consider the adversarial success as being able to validate a corrupted block or transaction, or forging a view during the Randomized Bootstrapping by finding $\signatureThreshold$ corrupted PoV validators, or imposing $\signatureThreshold$ corrupted view introducers within at most $\validatorThreshold$ trials, respectively. We evaluate the integrity of \textit{\lightchain} against the adversarial fraction (i.e., $\adv$) values of $0.16$, $0.33$, and $0.51$. $\adv = 0.16$ corresponds to the largest fraction of colluding hash power in the Bitcoin network \cite{cryptoeprint:2016:919}. Considering that each node in \textit{\lightchain} has a uniform chance of involvement in consensus, $\adv = 0.33$ is beyond the adversarial fraction threshold of BFT and PoS-based blockchains, e.g., Hyperledger \cite{androulaki2018hyperledger}. Likewise, $\adv = 0.51$ is beyond the adversarial power threshold of the PoW, e.g., Bitcoin \cite{nakamoto2008bitcoin}. As shown by Figure \ref{lightchain:fig_simulation_results}.a, for each of the selected adversarial fractions, there exists a certain value of $\validatorThreshold$ in which the success probability of adversarial peers in compromising the integrity of \textit{\lightchain} exponentially converges to zero with the growth of $\signatureThreshold$. Following Theorem \ref{lightchain:thm_integrity}, for $\adv = 0.16$ we obtain $\validatorThreshold \geq 10$ and $\signatureThreshold \geq 8$. Similarly, for $\adv = 0.33$, the theorem results in $\validatorThreshold \geq 20$ and $\signatureThreshold \geq 19$. Finally, for $\adv = 0.51$, the theorem results in $\validatorThreshold \geq 39$ and $\signatureThreshold \geq 37$. \textbf{This implies that the integrity of \textit{\lightchain} is preserved even when corrupted peers become the majority.} The values of $\validatorThreshold$ and $\signatureThreshold$ obtained form the security analysis for $\adv = 0.16, 0.33$, and $0.51$ are consistent with the results of Figure \ref{lightchain:fig_simulation_results}.a.

\textbf{Service Availability:} In contrast to Figure \ref{lightchain:fig_simulation_results}.a that solely considers the integrity of the system, Figure \ref{lightchain:fig_simulation_results}.b presents the integrity aspect of \textit{\lightchain} when its service availability is preserved. The validator thresholds of Figure \ref{lightchain:fig_simulation_results}.b are obtained by applying both Theorems \ref{lightchain:thm_data_availability} and \ref{lightchain:thm_service_availability}. Accordingly, we obtain $\validatorThreshold \geq 14$ and $\signatureThreshold \geq 11$ for $\adv = 0.16$, and $\validatorThreshold \geq 144$ and $\signatureThreshold \geq 80$ for $\adv = 0.33$. Theorems \ref{lightchain:thm_data_availability} and \ref{lightchain:thm_service_availability} are infeasible to satisfy together for $\adv > 0.5$. Hence, although there is no adversarial bound for the integrity of \textit{\lightchain} alone, its integrity under service availability is preserved for the adversarial fractions of less than $0.5$. This means that in contrast to the state-of-the-art blockchains such as Bitcoin \cite{nakamoto2008bitcoin}, Ethereum \cite{wood2014ethereum}, and Hyperledger \cite{androulaki2018hyperledger}, that are fully compromised once the adversarial fraction of nodes goes beyond their inherent threshold, our proposed \textit{\lightchain} system halts when the adversarial fraction of nodes goes beyond its configured operational parameters. This allows the blockchain community to notice the integrity risk and perform a decentralized bootstrapping of the system by discarding the nodes with suspicious behavior. We scope out the bootstrapping of \textit{\lightchain} after such a halt, and address it in our future works.

\textbf{Data Availability: }Figure \ref{lightchain:fig_simulation_results}.c illustrates the average availability of the validated blocks as $\signatureThreshold$ grows. By the blocks' availability, we mean the average number of available replicas for each block in the system at each time slot. The average is taken over $48$ hours of the simulation time. With $\signatureThreshold$ PoV validators for each block of \textit{\lightchain}, a block is replicated $\signatureThreshold + 1$ times in the system, i.e., $\signatureThreshold$ validators as well as the owner itself. As shown by Figure \ref{lightchain:fig_simulation_results}.c, the average availability of the blocks increases linearly with respect to $\signatureThreshold$. With uniform failure probability of $\ufp$, having $\signatureThreshold + 1$ replicas for a block or transaction results in $(\signatureThreshold + 1)\times(1-\ufp)$ available replicas in expectation. Hence, although following Theorem \ref{lightchain:thm_data_availability}, data availability of \textit{\lightchain} as a security measure is preserved with a high probability, as shown by Figure \ref{lightchain:fig_simulation_results}.c choosing $\signatureThreshold \geq \frac{1}{1-q} - 1$ results in an expected availability of at least one replica. In Figure \ref{lightchain:fig_simulation_results}.c, the number of replicas for $t = 1$ is about $1.6$ replicas on the average, and grows linearly as $\signatureThreshold$ increases.

\subsection{Cloud Platform Deployment}
To evaluate the performance of \textit{\lightchain} in practical setups, we implemented a proof-of-concept version of the \textit{\lightchain} node \cite{hassanzadeh2020containerized} over Skip Graph \cite{hassanzadeh2020skip}, and deployed a \textit{\lightchain} system of $1000$ nodes on the Google Cloud Platform. Due to the budget quota restrictions, each node in this deployment is running on low-power hardware with $2.2$ GHz of processing power and $3.9$ GB of memory. To avoid these low-power nodes crashing due to a high degree of concurrency, we serialized those operations of the nodes that require synchronized communication, i.e., resources getting blocked for a response or timeout. We performed deployments for the adversarial fraction of $0.16$ and $0.33$, each with the block sizes of $10$, $20$, and $40$ transactions. In each deployment, a node generates a transaction every second for a total of $1000$ transactions per node. The transaction time (i.e., generation, validation, and insertion in the Skip Graph overlay) is about $7$ and $82$ seconds for $\adv = 0.16$ and $\adv = 0.33$ on average, respectively. The approximate growth of $12$ times in the transaction time of $\adv = 0.33$ compared to $\adv = 0.16$ is due to the $10$ times growth in the number of validators of the former (i.e., $144$) compared to the latter (i.e., $14$). This linear growth behavior is due to the mentioned serialization of operations on the Google Cloud Platform that makes the operations including the request for transaction validation as well as the validation process of transactions done sequentially. A similar pattern is observed in the block formation time as shown by Figure \ref{lightchain:fig_poc_results}.a. The block formation time denotes the time that it takes for a node to collect a certain number of newly validated transactions into a block, generate the block metadata as specified in Section \ref{lightchain:sec_solution}, and request the validation of the block. Figure \ref{lightchain:fig_poc_results}.b shows the average local block validation time at each validator, which grows linearly with the number of transactions in the block. The difference between $\adv = 0.16$ and $\adv = 0.33$ in block validation time is the number of PoV signatures that each block validator verifies per transaction included in the block, which is a lighter operation than the network-based operations and is done with some degree of concurrency. 
Finally, Figure \ref{lightchain:fig_poc_results}.c presents a linear growth of the block size in KB as the number of transactions grows, which is expected as no size compression is done in the current proof-of-concept version of \textit{\lightchain}. The difference in the slope of $\adv = 0.33$ compared to $\adv = 0.16$ in Figure \ref{lightchain:fig_poc_results}.c is caused by their different number of required transaction validators, which their metadata and signatures are included in the blocks. At the proof-of-concept level, with serialized operations, and without any production-level optimizations, based on Figures \ref{lightchain:fig_poc_results}.a and \ref{lightchain:fig_poc_results}.b, \textit{\lightchain} generates an average of $1$ and $0.33$ Transactions Per Second (TPS) under the adversarial fractions of $\adv = 0.16$ and $\adv = 0.33$, which is slower than the production-grade transaction rate of blockchains like Ethereum with $15$ TPS \cite{schaffer2019performance}. \textbf{However, in contrast to best existing blockchains like Ethereum that are completely compromised on both integrity and service availability at $\adv = 0.33$, \textit{\lightchain} preserves both its integrity and service availability even at adversarial fractions of $\adv = 0.33$ and higher.} Additionally, our smaller scale \textit{\lightchain} simulations with concurrent and parallel operations show that \textit{\lightchain} is capable of approaching the average of about $60$ TPS. We leave optimizing the implementation of \textit{\lightchain} towards a production-grade one as our future work. 

\subsection{Asymptotic Analysis}
Having the system capacity of $\systemCapacity$ peers, a new peer joins the Skip Graph overlay using the original join protocol of the Skip Graph \cite{aspnes2007skip}, which has the message complexity of $O(\log{\systemCapacity})$. Generating and validating a new transaction or block has the message complexity of $O((\validatorThreshold + 1)\log{\systemCapacity})$, i.e., a maximum of $\validatorThreshold$ PoV Skip Graph searches, as well as an insertion in the overlay once it gets validated. 
This is simplified to $O(\log{\systemCapacity})$ considering $\validatorThreshold$ as a system-wide constant parameter. 
PoV of a single transaction or block is done by sending a validation result message, which has the message complexity of $O(1)$. Also, both generation and validation of a new transaction or block take a constant number of computational operations that are a function of the \textit{\lightchain}'s operational parameters (e.g., verifying $\signatureThreshold$ signatures from PoV validators), and takes $O(1)$ asymptotic time complexity. 
For the sake of \textit{\lightchain}'s Randomized-Bootstrapping, a peer needs to search for at most $\validatorThreshold$ view introducers within the overlay, which takes the message complexity of $O(\log{\systemCapacity})$. \textbf{Thus we conclude that the message complexity of \textit{\lightchain} is $O(\log{\systemCapacity})$ per operation.} Similarly, having $\blockCapacity$ blocks in the system, following the replication on validators policy of \textit{\lightchain}, \textbf{the expected storage complexity of each peer is $O(\frac{\blockCapacity}{\systemCapacity})$.} 
Based on the simulation results, with about $100$K generated blocks during the $48$ hours of simulation, the average standard deviation of the number of replicated blocks that each peer holds is about $0.033$, which corresponds to a uniform load distribution of blocks over the peers.

\section{Conclusion}
\label{lightchain:sec_conclusion}
To improve the communication and storage efficiency, and solve the convergence to centralization and consistency problems of the existing blockchain solutions, in this paper, we proposed \textit{\lightchain}, which is a novel blockchain architecture that operates over a Skip Graph-based P2P DHT overlay. In contrast to the existing blockchains that operate on epidemic data dissemination, \textit{\lightchain} provides addressable peers, blocks, and transactions within the network, which makes them efficiently accessible in an on-demand manner. Using \textit{\lightchain}, no peer is required to store the entire ledger. Rather each peer replicates a random subset of the blocks and transactions and answers other peer's queries on those.
\textit{\lightchain} is a fair blockchain as it considers a uniform chance for all the participating peers to be involved in the consensus protocol regardless of their influence in the system (e.g., hashing power or stake). To improve the consistency of the blockchain, \textit{\lightchain} governs a deterministic fork-resolving policy.

We analyzed \textit{\lightchain} both mathematically and experimentally regarding its operational parameters to achieve the security, efficiency, and availability. 
Having $\systemCapacity$ peers and $\blockCapacity$ blocks in the system, compared to the existing blockchains that require the storage and communication complexity of $O(\blockCapacity)$ and $O(\systemCapacity)$, respectively, \textit{\lightchain} requires  $O(\frac{\blockCapacity}{\systemCapacity})$ storage on each peer, and incurs the communication complexity of $O(\log{\systemCapacity})$ on generating a new transaction and block. 
As future work, we plan to extend our analysis to include game-theoretical aspects of \textit{\lightchain} (e.g., rewards and fees). 

\section*{Acknowledgement}
The authors thank Ali Utkan Şahin, Nazir Nayal, Shadi Sameh Hamdan, Mohammad Kefah Issa, Yüşa Ömer Altıntop, and Ece Tavona for their contributions to the implementation of \textit{\lightchain}.
We acknowledge the support of T\"{U}B\.{I}TAK (the Scientific and Technological Research Council of Turkey) project 119E088.

\bibliographystyle{IEEEtran}
\bibliography{references}

\begin{thebibliography}{10}
\providecommand{\url}[1]{#1}
\csname url@samestyle\endcsname
\providecommand{\newblock}{\relax}
\providecommand{\bibinfo}[2]{#2}
\providecommand{\BIBentrySTDinterwordspacing}{\spaceskip=0pt\relax}
\providecommand{\BIBentryALTinterwordstretchfactor}{4}
\providecommand{\BIBentryALTinterwordspacing}{\spaceskip=\fontdimen2\font plus
\BIBentryALTinterwordstretchfactor\fontdimen3\font minus
  \fontdimen4\font\relax}
\providecommand{\BIBforeignlanguage}[2]{{%
\expandafter\ifx\csname l@#1\endcsname\relax
\typeout{** WARNING: IEEEtran.bst: No hyphenation pattern has been}%
\typeout{** loaded for the language `#1'. Using the pattern for}%
\typeout{** the default language instead.}%
\else
\language=\csname l@#1\endcsname
\fi
#2}}
\providecommand{\BIBdecl}{\relax}
\BIBdecl

\bibitem{nakamoto2008bitcoin}
S.~Nakamoto, ``Bitcoin: A peer-to-peer electronic cash system,'' 2008.

\bibitem{reyna2018blockchain}
A.~Reyna, C.~Mart{\'\i}n, J.~Chen, E.~Soler, and M.~D{\'\i}az, ``On blockchain
  and its integration with iot. challenges and opportunities,'' \emph{Future
  Generation Computer Systems}, 2018.

\bibitem{khan2018iot}
M.~A. Khan and K.~Salah, ``Iot security: Review, blockchain solutions, and open
  challenges,'' \emph{Future Generation Computer Systems}, 2018.

\bibitem{ma2018blockchain}
Z.~Ma, M.~Jiang, H.~Gao, and Z.~Wang, ``Blockchain for digital rights
  management,'' \emph{Future Generation Computer Systems}, 2018.

\bibitem{mcconaghy2016bigchaindb}
T.~McConaghy, R.~Marques, A.~M{\"u}ller, D.~De~Jonghe, T.~McConaghy,
  G.~McMullen, R.~Henderson, S.~Bellemare, and A.~Granzotto, ``Bigchaindb: a
  scalable blockchain database,'' \emph{white paper, BigChainDB}, 2016.

\bibitem{jiang2017searchain}
P.~Jiang, F.~Guo, K.~Liang, J.~Lai, and Q.~Wen, ``Searchain: Blockchain-based
  private keyword search in decentralized storage,'' \emph{Future Generation
  Computer Systems}, 2017.

\bibitem{delgado2017fair}
S.~Delgado-Segura, C.~P{\'e}rez-Sola, G.~Navarro-Arribas, and
  J.~Herrera-Joancomart{\'\i}, ``A fair protocol for data trading based on
  bitcoin transactions,'' \emph{Future Generation Computer Systems}, 2017.

\bibitem{leng2018research}
K.~Leng, Y.~Bi, L.~Jing, H.-C. Fu, and I.~Van~Nieuwenhuyse, ``Research on
  agricultural supply chain system with double chain architecture based on
  blockchain technology,'' \emph{Future Generation Computer Systems}, 2018.

\bibitem{hassanzadeh2019decentralized}
Y.~Hassanzadeh-Nazarabadi, A.~K{\"u}p{\c{c}}{\"u}, and O.~Ozkasap,
  ``Decentralized utility-and locality-aware replication for heterogeneous
  dht-based p2p cloud storage systems,'' \emph{IEEE Transactions on Parallel
  and Distributed Systems}, vol.~31, no.~5, pp. 1183--1193, 2019.

\bibitem{kalodner2015empirical}
H.~A. Kalodner, M.~Carlsten, P.~Ellenbogen, J.~Bonneau, and A.~Narayanan, ``An
  empirical study of namecoin and lessons for decentralized namespace design.''
  in \emph{WEIS}, 2015.

\bibitem{croman2016scaling}
K.~Croman, C.~Decker, I.~Eyal, A.~E. Gencer, A.~Juels, A.~Kosba, A.~Miller,
  P.~Saxena, E.~Shi, E.~G. Sirer \emph{et~al.}, ``On scaling decentralized
  blockchains,'' in \emph{International Conference on Financial Cryptography
  and Data Security}.\hskip 1em plus 0.5em minus 0.4em\relax Springer, 2016.

\bibitem{eyal2016bitcoin}
I.~Eyal, A.~E. Gencer, E.~G. Sirer, and R.~Van~Renesse, ``Bitcoin-ng: A
  scalable blockchain protocol.'' in \emph{NSDI}, 2016.

\bibitem{nem2018}
G.~BloodyRookie and M.~Jaguar0625, ``Nem technical reference,''
  \emph{\url{https://nem.io/wp-content/themes/nem/files/NEM_techRef.pdf}},
  2018.

\bibitem{cryptoeprint:2016:919}
P.~Daian, R.~Pass, and E.~Shi, ``Snow white: Provably secure proofs of stake,''
  Cryptology ePrint Archive, 2016, \url{https://eprint.iacr.org/2016/919}.

\bibitem{kiayias2017ouroboros}
A.~Kiayias, A.~Russell, B.~David, and R.~Oliynykov, ``Ouroboros: A provably
  secure proof-of-stake blockchain protocol,'' in \emph{Annual International
  Cryptology Conference}.\hskip 1em plus 0.5em minus 0.4em\relax Springer,
  2017.

\bibitem{king2012ppcoin}
S.~King and S.~Nadal, ``Ppcoin: Peer-to-peer crypto-currency with
  proof-of-stake,''
  \emph{\url{https://peercoin.net/assets/paper/peercoin-paper.pdf}}, 2012.

\bibitem{bentov2014proof}
I.~Bentov, C.~Lee, A.~Mizrahi, and M.~Rosenfeld, ``proof of activity: Extending
  bitcoin's proof of work via proof of stake [extended abstract] y,'' \emph{ACM
  SIGMETRICS Performance Evaluation Review}, 2014.

\bibitem{neo2018}
\BIBentryALTinterwordspacing
NEO, ``Neo whitepaper,'' [Online; accessed 19-December-2018]. [Online].
  Available: \url{http://docs.neo.org/en-us/whitepaper.html}
\BIBentrySTDinterwordspacing

\bibitem{ontology2018}
\BIBentryALTinterwordspacing
ONT, ``Ontology whitepaper,'' [Online; accessed 23-December-2018]. [Online].
  Available: \url{https://dev-docs.ont.io}
\BIBentrySTDinterwordspacing

\bibitem{luu2016secure}
L.~Luu, V.~Narayanan, C.~Zheng, K.~Baweja, S.~Gilbert, and P.~Saxena, ``A
  secure sharding protocol for open blockchains,'' in \emph{SIGSAC}.\hskip 1em
  plus 0.5em minus 0.4em\relax ACM, 2016.

\bibitem{kokoris2018omniledger}
E.~Kokoris-Kogias, P.~Jovanovic, L.~Gasser, N.~Gailly, E.~Syta, and B.~Ford,
  ``Omniledger: A secure, scale-out, decentralized ledger via sharding,'' in
  \emph{IEEE SP}, 2018.

\bibitem{rocket2018snowflake}
T.~Rocket, ``Snowflake to avalanche: A novel metastable consensus protocol
  family for cryptocurrencies,'' 2018.

\bibitem{nikitin2017chainiac}
K.~Nikitin, E.~Kokoris-Kogias, P.~Jovanovic, N.~Gailly, L.~Gasser, I.~Khoffi,
  J.~Cappos, and B.~Ford, ``Chainiac: Proactive software-update transparency
  via collectively signed skipchains and verified builds,'' in \emph{USENIX
  Security 17}, 2017.

\bibitem{decker2016bitcoin}
C.~Decker, J.~Seidel, and R.~Wattenhofer, ``Bitcoin meets strong consistency,''
  in \emph{ICDCN}.\hskip 1em plus 0.5em minus 0.4em\relax ACM, 2016.

\bibitem{chepurnoy2016prunable}
A.~Chepurnoy, M.~Larangeira, and A.~Ojiganov, ``A prunable blockchain consensus
  protocol based on non-interactive proofs of past states retrievability.''
  \emph{arXiv preprint arXiv:1603.07926}, 2016.

\bibitem{otte2017trustchain}
P.~Otte, M.~de~Vos, and J.~Pouwelse, ``Trustchain: A sybil-resistant scalable
  blockchain,'' \emph{Future Generation Computer Systems}, 2017.

\bibitem{zamani2018rapidchain}
M.~Zamani, M.~Movahedi, and M.~Raykova, ``Rapidchain: Scaling blockchain via
  full sharding,'' in \emph{CCS}.\hskip 1em plus 0.5em minus 0.4em\relax ACM,
  2018.

\bibitem{androulaki2018hyperledger}
E.~Androulaki, A.~Barger, V.~Bortnikov, C.~Cachin, K.~Christidis, A.~De~Caro,
  D.~Enyeart, C.~Ferris, G.~Laventman, Y.~Manevich \emph{et~al.}, ``Hyperledger
  fabric: a distributed operating system for permissioned blockchains,'' in
  \emph{EuroSys}.\hskip 1em plus 0.5em minus 0.4em\relax ACM, 2018.

\bibitem{cachin2016architecture}
C.~Cachin, ``Architecture of the hyperledger blockchain fabric,'' in
  \emph{Workshop on distributed cryptocurrencies and consensus ledgers}, vol.
  310, 2016.

\bibitem{eyal2018majority}
I.~Eyal and E.~G. Sirer, ``Majority is not enough: Bitcoin mining is
  vulnerable,'' \emph{Communications of the ACM}, 2018.

\bibitem{zheng2017overview}
Z.~Zheng, S.~Xie, H.~Dai, X.~Chen, and H.~Wang, ``An overview of blockchain
  technology: Architecture, consensus, and future trends,'' in \emph{BigData
  congress}.\hskip 1em plus 0.5em minus 0.4em\relax IEEE, 2017.

\bibitem{zheng2018blockchain}
Z.~Zheng, S.~Xie, H.-N. Dai, X.~Chen, and H.~Wang, ``Blockchain challenges and
  opportunities: A survey,'' \emph{International Journal of Web and Grid
  Services}, 2018.

\bibitem{aspnes2007skip}
J.~Aspnes and G.~Shah, ``Skip graphs,'' \emph{ACM TALG}, 2007.

\bibitem{wust2017you}
K.~W{\"u}st and A.~Gervais, ``Do you need a blockchain?'' \emph{IACR Cryptology
  ePrint Archive}, 2017.

\bibitem{wood2014ethereum}
G.~Wood, ``Ethereum: A secure decentralised generalised transaction ledger,''
  \emph{Ethereum project yellow paper}, 2014.

\bibitem{douceur2002sybil}
J.~R. Douceur, ``The sybil attack,'' in \emph{International workshop on
  peer-to-peer systems}.\hskip 1em plus 0.5em minus 0.4em\relax Springer, 2002.

\bibitem{hassanzadeh2018decentralized}
Y.~Hassanzadeh-Nazarabadi, A.~K{\"u}p{\c{c}}{\"u}, and {\"O}.~{\"O}zkasap,
  ``Decentralized and locality aware replication method for dht-based p2p
  storage systems,'' \emph{Future Generation Computer Systems}, 2018.

\bibitem{hassanzadeh2016awake}
------, ``Awake: decentralized and availability aware replication for p2p cloud
  storage,'' in \emph{2016 IEEE International Conference on Smart Cloud
  (SmartCloud)}.\hskip 1em plus 0.5em minus 0.4em\relax IEEE, 2016, pp.
  289--294.

\bibitem{hassanzadeh2020skipsim}
Y.~Hassanzadeh-Nazarabadi, A.~U. {\c{S}}ahin, {\"O}.~{\"O}zkasap, and
  A.~K{\"u}p{\c{c}}{\"u}, ``Skipsim: Scalable skip graph simulator,'' in
  \emph{2020 IEEE International Conference on Blockchain and Cryptocurrency
  (ICBC)}.\hskip 1em plus 0.5em minus 0.4em\relax IEEE, 2020, pp. 1--2.

\bibitem{lightchain-container}
Y.~Hassanzadeh-Nazarabadi, N.~Nayal, S.~S. Hamdan, {\"O}.~{\"O}zkasap, and
  A.~K{\"u}p{\c{c}}{\"u}, ``A containerized proof-of-concept implementation of
  lightchain system,'' in \emph{2020 IEEE International Conference on
  Blockchain and Cryptocurrency (ICBC)}.\hskip 1em plus 0.5em minus 0.4em\relax
  IEEE, 2020, pp. 1--2.

\bibitem{hassanzadeh2017elats}
Y.~Hassanzadeh-Nazarabadi and {\"O}.~{\"O}zkasap, ``Elats: Energy and locality
  aware aggregation tree for skip graph,'' in \emph{2017 IEEE International
  Black Sea Conference on Communications and Networking (BlackSeaCom)}.\hskip
  1em plus 0.5em minus 0.4em\relax IEEE, 2017, pp. 1--5.

\bibitem{jakobsson1999proofs}
M.~Jakobsson and A.~Juels, ``Proofs of work and bread pudding protocols,'' in
  \emph{Secure Information Networks}.\hskip 1em plus 0.5em minus 0.4em\relax
  Springer, 1999.

\bibitem{kraft2016difficulty}
D.~Kraft, ``Difficulty control for blockchain-based consensus systems,''
  \emph{Peer-to-Peer Networking and Applications}, 2016.

\bibitem{laurie2004proof}
B.~Laurie and R.~Clayton, ``Proof-of-work proves not to work; version 0.2,'' in
  \emph{Workshop on Economics and Information, Security}, 2004.

\bibitem{vukolic2015quest}
M.~Vukoli{\'c}, ``The quest for scalable blockchain fabric: Proof-of-work vs.
  bft replication,'' in \emph{International workshop on open problems in
  network security}.\hskip 1em plus 0.5em minus 0.4em\relax Springer, 2015.

\bibitem{o2014bitcoin}
K.~O'Dwyer and D.~Malone, ``Bitcoin mining and its energy footprint,'' in
  \emph{IET}.\hskip 1em plus 0.5em minus 0.4em\relax The Institution of
  Engineering \& Technology, 2014.

\bibitem{buterin2017casper}
V.~Buterin and V.~Griffith, ``Casper the friendly finality gadget,''
  \emph{arXiv preprint arXiv:1710.09437}, 2017.

\bibitem{lamport1978implementation}
L.~Lamport, ``The implementation of reliable distributed multiprocess
  systems,'' \emph{Computer Network}, 1978.

\bibitem{dwork1992pricing}
C.~Dwork and M.~Naor, ``Pricing via processing or combatting junk mail,'' in
  \emph{Annual International Cryptology Conference}.\hskip 1em plus 0.5em minus
  0.4em\relax Springer, 1992.

\bibitem{liu2006proof}
D.~Liu and L.~J. Camp, ``Proof of work can work,'' \emph{WEIS}, 2006.

\bibitem{poelstra2014distributed}
A.~Poelstra \emph{et~al.}, ``Distributed consensus from proof of stake is
  impossible,'' \emph{URL: https://download. wpsoftware. net/bitcoin/old-pos.
  pdf}, 2014.

\bibitem{pease1980reaching}
M.~Pease, R.~Shostak, and L.~Lamport, ``Reaching agreement in the presence of
  faults,'' \emph{JACM}, 1980.

\bibitem{schwartz2014ripple}
D.~Schwartz, N.~Youngs, A.~Britto \emph{et~al.}, ``The ripple protocol
  consensus algorithm,'' \emph{Ripple Labs Inc White Paper}, 2014.

\bibitem{kwon2014tendermint}
J.~Kwon, ``Tendermint: Consensus without mining,'' \emph{Retrieved May},
  vol.~18, p. 2017, 2014.

\bibitem{lamport1998part}
L.~Lamport \emph{et~al.}, ``The part-time parliament,'' \emph{ACM TOCS}, 1998.

\bibitem{harris2018holochain}
E.~Harris-Braun, N.~Luck, and A.~Brock, ``Holochain-scalable agentcentric
  distributed computing,'' \emph{Alpha}, 2018.

\bibitem{cassandra2014apache}
A.~Cassandra, ``Apache cassandra,'' \emph{Website. Available online at
  http://planetcassandra. org/what-is-apache-cassandra}, 2014.

\bibitem{back2014enabling}
A.~Back, M.~Corallo, L.~Dashjr, M.~Friedenbach, G.~Maxwell, A.~Miller,
  A.~Poelstra, J.~Tim{\'o}n, and P.~Wuille, ``Enabling blockchain innovations
  with pegged sidechains,'' \emph{\url{http://www.opensciencereview.
  com/papers/123/enablingblockchain-innovations-with-pegged-sidechains}}, 2014.

\bibitem{pugh1990skip}
W.~Pugh, ``Skip lists: a probabilistic alternative to balanced trees,''
  \emph{Communications of the ACM}, 1990.

\bibitem{zyskind2015decentralizing}
G.~Zyskind, O.~Nathan \emph{et~al.}, ``Decentralizing privacy: Using blockchain
  to protect personal data,'' in \emph{Security and Privacy Workshops}.\hskip
  1em plus 0.5em minus 0.4em\relax IEEE, 2015.

\bibitem{maymounkov2002kademlia}
P.~Maymounkov and D.~Mazieres, ``Kademlia: A peer-to-peer information system
  based on the xor metric,'' in \emph{International Workshop on Peer-to-Peer
  Systems}.\hskip 1em plus 0.5em minus 0.4em\relax Springer, 2002.

\bibitem{hassanzadeh2015locality}
Y.~Hassanzadeh-Nazarabadi, A.~K{\"u}p{\c{c}}{\"u}, and {\"O}.~{\"O}zkasap,
  ``Locality aware skip graph,'' in \emph{IEEE ICDCSW, 2015}.

\bibitem{hassanzadeh2016laras}
------, ``Laras: Locality aware replication algorithm for the skip graph,'' in
  \emph{IEEE NOMS}, 2016.

\bibitem{etemad2015efficient}
M.~Etemad and A.~K{\"u}p{\c{c}}{\"u}, ``Efficient key authentication service
  for secure end-to-end communications,'' in \emph{ProvSec}.\hskip 1em plus
  0.5em minus 0.4em\relax Springer, 2015.

\bibitem{stutzbach2006understanding}
D.~Stutzbach and R.~Rejaie, ``Understanding churn in peer-to-peer networks,''
  in \emph{SIGCOMM}.\hskip 1em plus 0.5em minus 0.4em\relax ACM, 2006.

\bibitem{jacob2014skip+}
R.~Jacob, A.~Richa, C.~Scheideler, S.~Schmid, and H.~T{\"a}ubig, ``Skip+: A
  self-stabilizing skip graph,'' \emph{JACM}, 2014.

\bibitem{hassanzadeh2019interlaced}
Y.~Hassanzadeh-Nazarabadi, A.~K{\"u}p{\c{c}}{\"u}, and {\"O}.~{\"O}zkasap,
  ``Interlaced: Fully decentralized churn stabilization for skip graph-based
  dhts,'' \emph{arXiv preprint arXiv:1903.07289}, 2019.

\bibitem{tanenbaum2007distributed}
A.~S. Tanenbaum and M.~Van~Steen, \emph{Distributed systems: principles and
  paradigms}.\hskip 1em plus 0.5em minus 0.4em\relax Prentice-Hall, 2007.

\bibitem{goodrich2011introduction}
M.~T. Goodrich and R.~Tamassia, \emph{Introduction to computer security}.\hskip
  1em plus 0.5em minus 0.4em\relax Pearson, 2011.

\bibitem{boshrooyeh2017guard}
S.~T. Boshrooyeh and O.~Ozkasap, ``Guard: Secure routing in skip graph,'' in
  \emph{IFIP Networking}.\hskip 1em plus 0.5em minus 0.4em\relax IEEE, 2017.

\bibitem{taheri2020proof}
S.~Taheri-Boshrooyeh, A.~U. {\c{S}}ahin, Y.~Hassanzadeh-Nazarabadi, and
  {\"O}.~{\"O}zkasap, ``A proof-of-concept implementation of guard secure
  routing protocol,'' in \emph{2020 International Symposium on Reliable
  Distributed Systems (SRDS)}.\hskip 1em plus 0.5em minus 0.4em\relax IEEE,
  2020, pp. 332--334.

\bibitem{stoica2001chord}
I.~Stoica, R.~Morris, D.~Karger, M.~F. Kaashoek, and H.~Balakrishnan, ``Chord:
  A scalable peer-to-peer lookup service for internet applications,'' \emph{ACM
  SIGCOMM Computer Communication Review}, 2001.

\bibitem{rowstron2001pastry}
A.~Rowstron and P.~Druschel, ``Pastry: Scalable, decentralized object location,
  and routing for large-scale peer-to-peer systems,'' in \emph{IFIP/ACM
  International Conference on Distributed Systems Platforms and Open
  Distributed Processing}.\hskip 1em plus 0.5em minus 0.4em\relax Springer,
  2001.

\bibitem{li2020survey}
X.~Li, P.~Jiang, T.~Chen, X.~Luo, and Q.~Wen, ``A survey on the security of
  blockchain systems,'' \emph{FGCS, Elsevier}, 2020.

\bibitem{katz2014introduction}
J.~Katz and Y.~Lindell, \emph{Introduction to modern cryptography}.\hskip 1em
  plus 0.5em minus 0.4em\relax CRC press, 2014.

\bibitem{harkness1965properties}
W.~L. Harkness, ``Properties of the extended hypergeometric distribution,''
  \emph{The Annals of Mathematical Statistics}, 1965.

\bibitem{bertsekas2002introduction}
D.~P. Bertsekas and J.~N. Tsitsiklis, \emph{Introduction to probability}.\hskip
  1em plus 0.5em minus 0.4em\relax Athena Scientific Belmont, MA, 2002.

\bibitem{imtiaz2019churn}
M.~A. Imtiaz, D.~Starobinski, A.~Trachtenberg, and N.~Younis, ``Churn in the
  bitcoin network: Characterization and impact,'' in \emph{ICBC}.\hskip 1em
  plus 0.5em minus 0.4em\relax IEEE, 2019.

\bibitem{bitnodes}
``Global bitcoin nodes distribution,'' \url{https://bitnodes.earn.com/},
  accessed: 24-09-2018.

\bibitem{blockchaincharts}
``Blockchain charts,'' \url{https://www.blockchain.com/en/charts}, accessed:
  24-09-2018.

\bibitem{hassanzadeh2020containerized}
Y.~Hassanzadeh-Nazarabadi, N.~Nayal, S.~S. Hamdan, {\"O}.~{\"O}zkasap, and
  A.~K{\"u}p{\c{c}}{\"u}, ``A containerized proof-of-concept implementation of
  lightchain system,'' in \emph{ICBC}.\hskip 1em plus 0.5em minus 0.4em\relax
  IEEE, 2020.

\bibitem{hassanzadeh2020skip}
Y.~Hassanzadeh-Nazarabadi, N.~Nayal, S.~S. Hamdan, A.~U. {\c{S}}ahin,
  {\"O}.~{\"O}zkasap, and A.~K{\"u}p{\c{c}}{\"u}, ``Skip graph middleware
  implementation,'' in \emph{SRDS}.\hskip 1em plus 0.5em minus 0.4em\relax
  IEEE, 2020.

\bibitem{schaffer2019performance}
M.~Sch{\"a}ffer, M.~Di~Angelo, and G.~Salzer, ``Performance and scalability of
  private ethereum blockchains,'' in \emph{International Conference on Business
  Process Management}.\hskip 1em plus 0.5em minus 0.4em\relax Springer, 2019,
  pp. 103--118.

\end{thebibliography}
\newpage
\appendix
\section{\lightchain's algorithms details}
\label{lightchain:sec_appendix}
In this appendix, we represent the detailed descriptions of the \textit{\lightchain}'s algorithms in a bottom-up manner, i.e., we first show the basic low-level algorithms that act as the building blocks of the high-level ones, and then move to the explanation of the high-level algorithms that operate on the top of those building blocks. 
The Verify algorithm (Algorithm \ref{lightchain:alg_verify}) verifies the authenticity of the input search proof, and generates a misbehavior transaction upon receiving an unauthenticated search proof (see Section \ref{lightchain:subsection_auditing} for more details). The TXB-Generation algorithm (Algorithm \ref{lightchain:txblk_gen}) is called whenever a peer aims to generate a transaction or block, and it uses the Verify algorithm as a sub-routine. Algorithms \ref{lightchain:alg_issound}, \ref{lightchain:alg_iscorrect}, and \ref{lightchain:alg_isauthenticated} are sub-routines that evaluate the soundness, correctness, and authenticity of a transaction as described in Section \ref{lightchain:tx_validation}, respectively. Algorithm \ref{lightchain:alg_isauthenticated} also evaluates the authenticity of blocks, which is done in an identical manner as for the transactions. 

Using these building blocks, the Audit algorithm (Algorithm \ref{lightchain:alg_audit}) evaluates the validity of a given transaction or block based on the view of the peer that invokes it, and generates a misbehavior transaction on receiving an invalid transaction or block (see Section \ref{lightchain:subsection_auditing} for further details). The Audit algorithm is used as a sub-routine in the ViewUpdate algorithm (Algorithm \ref{lightchain:alg_viewupdate}), which performs randomized bootstrapping for a new peer that joins the system (see Section \ref{lightchain:subsection_bootstrap}) as well as updating the view of the existing peers in the system. For an already joined peer to the system, a single call to ViewUpdate updates the view of the peer towards the tail of the blockchain by one block, e.g., a peer with a view that is three blocks behind the current tail of the blockchain needs to invoke ViewUpdate three times to reach the current tail of the ledger. \textit{\lightchain} peers invoke ViewUpdate periodically to update their view towards the blockchain. The frequency of  ViewUpdate execution is application dependent. ViewUpdate audits the newly generated transactions and blocks against the misbehavior, and generates a new block by invoking the TXB-Generation algorithm upon collecting \txnum newly generated transactions. Algorithm \ref{lightchain:alg_hasbalance} evaluates the balance compliance of a transaction owner to cover the routing and validation fees based on the view of the PoV validator that invokes it. Finally, Algorithm \ref{lightchain:alg_pov} is executed by a PoV validator peer and represents the PoV validation procedure of a transaction or block. 

\begin{algorithm}
\KwIn{proof of search $search\_proof$, numerical ID of the peer $numID$, routing table of the peer $table$, local view of the peer on blockchain's tail $c\_tail$}
\KwOut{boolean $result$}
\tcp{Validate the search proof based on the validation mechanism of underlying authenticated search proof}
\uIf{$search\_proof$ is authenticated}
{
    $result = true$\;
}
\uElse
{
    $result = false$\;
    \tcp{Misbehavior detection (see Section \ref{lightchain:subsection_auditing})}
    find the guilty node and report it in $cont$\;
    \tcp{Generate a misbehavior transaction}
    TXB-Generation($numID$, $table$, $c\_tail$, $cont$)\;
}

\caption{Verify}
\label{lightchain:alg_verify}
\end{algorithm}
\begin{algorithm}
\KwIn{identifier of the owner $owner$, routing table of the owner $table$, previous pointer $\prev$, contributions $cont$ \textbf{or} transaction set $\mathcal{S}$}
\KwOut{a new transaction \textbf{or} block $txblk$}
\While{i $\in [1,\validatorThreshold]$}
 {
   \tcp{Search for the $i^{th}$ validator within overlay}
   $v_{i} = H(\prev||owner||cont/\mathcal{S}||i)$\;
   $search\_proof_{i} = $ searchForNumericalID($v_{i}$, $table$)\;
  \If{Verify($search\_proof_{i},\,$ $owner,\,$ $table,\,$ $\prev$)}
  {
    add $search\_proof_{i}$ to $search\_proof$\;
  }
  increase $i$\;
 }
 include $\prev$ and $search\_proof$ into $txblk$\;
 include $cont/\mathcal{S}$ into $txblk$\;
 compute the hash and include it in $txblk.h$\;
 sign $txblk.h$\;
 include signature in $txblk.\sigma$\;
 send $txblk$ for validation to the validators\;
 obtain the validation signatures from validators\;
 \If{$\signatureThreshold$ validators signatures on $txblk$ obtained}
 {
    include validators signature in $txblk.\sigma$\;
    insert $txblk$ into overlay as a node\;
    \If{$txblk$ is a block}
    {
            \If{$txblk$ is knocked-out in a fork}
            {   
               drop $txblk$ from the overlay\;
               terminate\;
            }
            \uElse
            {
                \tcp{Adding transaction pointers for direct state retrieval (See Section \ref{lightchain:subsection_direct_access} for more details)}
                \For{$tx \in txblk.\mathcal{S}$} 
                {
                    insert a transaction pointer to $tx.owner$ into the overlay\;
                }
            }
    }
    \If{$txblk$ is a transaction}
    {
        \While{$txblk$ is not included in a block}
        {wait\;}
        drop $txblk$ from the overlay\;
    }
    
 }
 
\caption{TXB-Generation}
\label{lightchain:txblk_gen}
\end{algorithm}

\begin{algorithm}
\KwIn{transaction $tx$, view of the auditor/validator $view$}
\KwOut{boolean $result$}
\tcp{Retrieve the address of the last block of that contains the most recent transaction of the $tx$'s owner}
($lastblk$, $state$, $balance$) = $view.get(tx.owner)$\; 
\uIf{$tx.\prev$ points to a predecessor of $lastblk$}
{  
    \tcp{$tx$ is not sound as it violates the total ordering among the transactions of $tx.owner$}
    $result = false$\;
}
\uElse
{
    {$result = true$\;}
}
\caption{isSound}
\label{lightchain:alg_issound}
\end{algorithm}
\begin{algorithm}
\KwIn{transaction $tx$, view of the auditor/validator $view$}
\KwOut{boolean $result$}
\tcp{Retrieve the state of the transaction owner from the view of the auditor/validator peer}
($lastblk$, $state$, $balance$) = $view.get(tx.owner)$\;
\uIf{$tx.cont$ contains a valid transition of $state$}
{  
    $result = true$\;
}
\uElse
{
    {$result = false$\;}
}
\caption{isCorrect}
\label{lightchain:alg_iscorrect}
\end{algorithm}
\begin{algorithm}
\KwIn{transaction/block $txblk$, numerical ID of the peer $numID$, routing table of the peer $table$, local view of the peer on blockchain's tail $c\_tail$}
\KwOut{boolean $result$}
\tcp{Result is initially true and is set to false upon detection of an authenticity violation}
$result = true$\;
\uIf{$txblk.h \neq  H(prev||owner||cont/\mathcal{S}||search\_proof)$}
{  
    \tcp{Invalid hash value}
    $result = false$\;
}
\uElseIf{valid signature of $txblk.owner$ on $txblk.h \not \in txblk.\sigma$}
{
    \tcp{Missing a valid signature of the owner}
    $result = false$\;
}
\uElse
{
    \tcp{Check the number of PoV validators that signed the transaction/block}
    $validatorCounter = 0$\;
    \For{i $\in [1,\validatorThreshold]$}
    {
     $v_{i} = H(\prev||owner||cont/\mathcal{S}||i)$\;
     \If{$search\_proof_{i} \in search\_proof  \land$  $Verify(search\_proof_{i},\,$ $numID,\,$ $table,\, c\_tail) \land$ 
     signature of $i^{th}$ validator on $txblk.h \in$ $txblk.\sigma$}
     {
        $validatorCounter++$\;
        \If{$validatorCounter == \signatureThreshold$}
        {
            break\;
        }
     }
    }
     \If{$validatorCounter < \signatureThreshold$}
     {
        $result = false\;$
     }
}
\caption{isAuthenticated}
\label{lightchain:alg_isauthenticated}
\end{algorithm}
\begin{algorithm}
\KwIn{transaction/block $txblk$, the auditor view $view$, numerical ID of the auditor $numID$, routing table of the auditor $table$, local view of the auditor on blockchain's tail $c\_tail$}
\KwOut{boolean $result$}
\uIf{$txblk$ is a transaction}
{
    
    {$result = isSound(txblk, view) \land isAuthenticated(txblk,\, numID,\, table,\, c\_tail)$\;}
}
\uElse
{ 
 \tcp{$txblk$ is a block}
 \If{$isAuthenticated(txblk,\, numID,\, table,\, c\_tail)$}
 {
    \tcp{The result is initially set to true for an authenticated block, but its final value depends on the soundness and authenticity of each individual transaction inside the block}
    $result = true$\;
    \For{transaction $tx \in txblk.\mathcal{S}$} 
    {
         \If{$\neg isSound(tx, view) \lor \neg isAuthenticated(tx,\, numID,\, table,\, c\_tail)$}
          {
                $result$ = $false$\;
                break\;
          }
    } 
 }
 
 \If{$result == false$}
 {
    \tcp{Misbehavior detection (see Section \ref{lightchain:subsection_auditing})}
    find the guilty node and report it in $cont$\;
    \tcp{Generate a misbehavior transaction}
    TXB-Generation($numID$, $table$, $c\_tail$, $cont$)\;
 }
}
\caption{Audit}
\label{lightchain:alg_audit}
\end{algorithm}
\begin{algorithm}
\KwIn{numerical ID of the peer $numID$, routing table of the peer $table$, local view of the peer on blockchain's tail $c\_tail$, view of the peer $view$, set of peer's collected new transactions $\mathcal{S}$}
\KwOut{updated $view$, updated $\mathcal{S}$}

\eIf{$view$ is empty}
{  
    \tcp{An empty view corresponds to a new node that requires randomized bootstrapping to create its view}
    \While{i $\in [1,\validatorThreshold]$ $\land$ less than $\signatureThreshold$ consistent views obtained}
    {
       \tcp{Find the $i^{th}$ view introducer}
       $view\_intro_{i} = H(numID||i)$ \;
       $search\_proof_{i} = $ searchForNumericalID($view\_intro_{i},\, table$)\;
        \If{Verify($search\_proof_{i},\,$ $numID,\,$ $table,\,$ $c\_tail)$}
          {
            contact $i^{th}$ view introducer and obtain its view\;
          }
    }
    \If{$\signatureThreshold$ consistent views obtained}
    {
        update $view$\; 
    }
       
}
{
  \tcp{Update the view towards the current tail of the blockchain using a search for name ID of the local view of the current tail within the overlay}
   $search\_proof_{tail} = $ searchForNameID($c\_tail,\,
  \{table\}$)\;
  \If{$Verify(search\_proof_{tail},\,$ $numID,\,$ $table,\,$ $c\_tail$) }
  {
     \uIf{set of new transactions $\{TX\} \in search\_proof_{tail}$}
     {

        add $tx \in \{TX\}$ with $Audit(tx,\, view,\, numID,\, table,\, c\_tail)$ == $true$ to $\mathcal{S}$\;
        \If{$\txnum$ new transactions are in $\mathcal{S}$}
        {
          \tcp{Generate a block out of the collected transactions}
          TXB-Generation($numID,\,$ $table,\,$  $c\_tail,\,$ $\mathcal{S}$)\;
        }
     }
     \uElseIf{new block(s) found}
     {
        \If{there is a fork}
        {follow the block $blk$ with minimum hash value\;}
         \tcp{Misbehavior verification}
         \If{$Audit(blk,\, view,\, numID,\, table,\, c\_tail)$}
         {
            update the $view$ based on the new block\;
            $c\_tail = blk$\; 
            \tcp{Dropping existing transaction pointers (See Section \ref{lightchain:subsection_direct_access} for more details)}
            \For{$tx \in blk.\mathcal{S}$} 
            {
                \If{holding a transaction pointer to $tx.owner$}
                {drop the transaction pointer\;}
            }
         }
     }
  }
}
\caption{ViewUpdate}
\label{lightchain:alg_viewupdate}
\end{algorithm}
\begin{algorithm}
\KwIn{transaction $tx$, view of the validator $view$}
\KwOut{boolean $result$}
\tcp{Retrieve the state of the transaction owner from the view of the validator peer}
($lastblk$, $state$, $balance$) = $view.get(tx.owner)$\;
\uIf{owner has enough $balance$ to cover the routing and validation fees}
{  
    $result = true$\;
}
\uElse
{
    {$result = false$\;}
}
\caption{hasBalanceCompliance}
\label{lightchain:alg_hasbalance}
\end{algorithm}

\begin{algorithm}
\KwIn{transaction/block $txblk$, numerical ID of the validator peer $numID$, view of the validator $view$, local view of the validator on blockchain's tail $c\_tail$}
\KwOut{message $msg$}

\eIf{$txblk$ is a transaction}
{
    \tcp{The validation of a transaction}
    \tcp{Check the transaction's soundness, correctness, authenticity, and the balance compliance of its owner}
    $msg = isSound(txblk,\, view)\land  isCorrect(txblk,\, view) \land isAuthenticated(txblk,\, numID,\, table,\, c\_tail) \land hasBalanceCompliance(txblk, view)$\;
    }
    {
        \tcp{The validation of a block}
        \tcp{Check the authenticity and consistency of the block}
        \eIf{$isAuthenticated(txblk,\, numID,\, table,\, c\_tail) \land txblk.prev == c\_tail$}
        {
            \tcp{Check the soundness and authenticity of each individual transaction in the block}
            \For{$tx \in txblk.\mathcal{S}$} 
            {
                $msg = isSound(tx,\, view) \land isAuthenticated(tx,\, numID,\, table,\,$ $c\_tail)$\;
                \If{more than one transaction from $tx.owner \in txblk.\mathcal{S}$}
                {
                  $msg = false$\;
                }
                \If{$msg == false$}
                {
                    break\;
                }
            } 
        }
        {
           \tcp{A consistency or authenticity violation found in the block}
           $msg$ = $false\;$ 
        }
}
\If{$msg == true$}
{
    $msg$ = signature on $txblk.h$ by the validator's signing key\;
    \tcp{Holding a replica of the validated transaction or block}
    insert $txblk$ into the overlay as a node\;
    \If{$txblk$ is a block}
    {
            \If{$txblk$ is knocked-out in a fork}
            {   
               drop $txblk$ from the overlay\;
               terminate\;
            }
            \uElse
            {
                \tcp{Adding transaction pointers for direct state retrieval (See Section \ref{lightchain:subsection_direct_access} for more details)}
                \For{$tx \in txblk.\mathcal{S}$} 
                {
                    insert a transaction pointer to $tx.owner$ into the overlay\;
                }
            }
    }
    \If{$txblk$ is a transaction}
    {
        \While{$txblk$ is not included in a block}
        {wait\;}
        drop $txblk$ from the overlay;
    }
}
send $msg$ to $txblk.owner$\;
\caption{Proof-of-Validation (PoV)}
\label{lightchain:alg_pov}
\end{algorithm}
\end{document}